\documentclass[aip,reprint,pop]{revtex4-1}

\usepackage{graphicx}

\newcommand\coline{(Color online)\ }

\newcommand\sech{\ensuremath{\mathrm{sech}}}
 
\newcommand\dif{\ensuremath{\mathrm{d}}}

\begin{document} 

\title{Resistive magnetohydrodynamic simulations of {X}-line retreat
  during magnetic reconnection}

\author{N. A. Murphy}
\email[]{namurphy@cfa.harvard.edu}
\affiliation{Harvard-Smithsonian Center for Astrophysics, Cambridge,
  Massachusetts 02138, USA}

\date{\today} 

\keywords{Magnetic reconnection, plasma simulation, plasma
  magnetohydrodynamics}

\begin{abstract}
To investigate the impact of current sheet motion on the reconnection
process, we perform resistive magnetohydrodynamic (MHD) simulations of
two closely located reconnection sites which move apart from each
other as reconnection develops.
This simulation develops less quickly than an otherwise equivalent
single perturbation simulation but eventually exhibits a higher
reconnection rate.
The unobstructed outflow jets are faster and longer than the outflow
jets directed towards the magnetic island that forms between the two
current sheets.
The X-line and flow stagnation point are located near the trailing end
of each current sheet very close to the obstructed exit.  
The speed of X-line retreat ranges from {$\sim$}$0.02$--$0.06$ while
the speed of stagnation point retreat ranges from
{$\sim$}$0.03$--$0.07$, in units of the initial upstream Alfv\'en
velocity.
Early in time, the flow stagnation point is located closer to the
center of the current sheet than the X-line, but later on the relative
positions of these two points switch.
Consequently, late in time there is significant plasma flow across the
X-line in the opposite direction of X-line retreat.
Throughout the simulation, the velocity \emph{at} the X-line does not
equal the velocity \emph{of} the X-line.
Motivated by these results, an expression for the rate of X-line
retreat is derived in terms of local parameters at the X-point.
This expression shows that X-line retreat is due to both advection by
the bulk plasma flow and diffusion of the normal component of the
magnetic field.
\end{abstract}
\pacs{52.35.Vd,52.65.-y} 

\maketitle 

\section{INTRODUCTION\label{introduction}}

Most simulations and theories of magnetic reconnection operate under
the assumption that the current sheet is roughly stationary with
respect to the ambient plasma.  However, there are many situations in
nature and the laboratory where current sheet motion is important.
Recently, a model was developed to describe steady magnetic
reconnection with asymmetric outflow in a high aspect ratio current
sheet.\cite{murphy:asym} While the assumption of time-independence was
needed to make analytic progress, such an assumption precludes
time-dependent effects such as current sheet motion.  This paper
addresses this issue by presenting resistive magnetohydrodynamic (MHD)
simulations of two X-lines which start in close proximity to each
other and move apart as reconnection develops.

The near-Earth neutral line model\cite{baker:1996} predicts that the
magnetotail X-line retreats in the tailward direction during the
recovery phase of magnetospheric substorms.  Such behavior is commonly
observed during \emph{in situ} measurements in the Earth's
magnetotail.\cite{runov:2003,eastwood:2010} In a recent statistical
study of diffusion region crossings by Cluster, tailward moving
X-lines were observed {$\sim$}{$4$} times more frequently than
earthward moving X-lines.\cite{eastwood:2010}
Despite the common occurrence of current sheet motion, it is standard
practice to compare \emph{in situ} measurements of diffusion region
crossings to particle-in-cell (PIC) simulations of roughly stationary
reconnection layers.  Global simulations of the magnetotail do allow
the diffusion region to move, and tailward motion of the predominant
X-line is commonly observed.\cite{birn:1996, ohtani:2004,
kuznetsova:2007, zhu:2009} Previous research has also considered the
effects of current sheet motion on reconnection slow mode shock
structure.\cite{owen:1987:451, *owen:1987:467, kiehas:2007,
kiehas:2009}

Flux rope models of coronal mass ejections (CMEs) typically predict
the formation of a current sheet between the flare site and the
ejected plasmoid.\cite{kopp:pneuman:1976, forbes:acton:1996,
  linforbes:2000} Features identified as current sheets
have been observed for an increasing number of events (e.g., Refs.\
\onlinecite{2002ApJ...575.1116C, 2008ApJ...686.1372C,
2010ApJ...708.1135S, savage:2010}).
During these events both the lower and upper boundaries of the current
sheet are thought to rise with time.\cite{linker:2003,reeves:2010} Of
particular interest are the locations of the predominant X-line and
flow stagnation point.  A recent analysis of the current sheet behind
a slow CME on 2008 April 9 showed upflowing features above and
downflowing features below a height of $\sim${$0.25$} solar
radii.\cite{savage:2010} In addition, a potential field model of the
pre-CME active region showed a pre-existing X-line at about that
height.  Because the CME current sheet associated with this event
extended beyond several solar radii, this evidence suggests that the
predominant X-line and flow stagnation point are both located near the
base of the current sheet.  

Current sheet motion and asymmetric outflow reconnection occur in
laboratory plasma devices involving the merging of spheromaks and
toroidal plasma configurations where the outflow is aligned with the
radial direction, even though the range of motion is limited by the
boundary conditions of the experiment.  
Counter-helicity spheromak merging results from the Magnetic
Reconnection Experiment (MRX)\cite{yamada:mrx} show the X-line being
pulled towards one end of the current sheet because of the Hall
effect, resulting in asymmetric outflow.\cite{inomoto:counter,
murphy:mrx}
During spheromak merging experiments at TS-3/4,\cite{ono:1993} current
sheet ejection often leads to faster reconnection and the X-line being
located towards one end of the current sheet.\cite{ono:1997}

Current sheet motion also occurs when both a guide field and a density
gradient across the current layer are present.\cite{rogers:1995,
swisdak:diamagnetic, phan:2010} Relevant configurations include the dayside
magnetopause and tokamaks.  In these situations the plasma pressure
gradient in the inflow direction leads to diamagnetic drifting of the
reconnection layer.  When the diamagnetic drift velocity becomes
comparable to the Alfv\'en speed, reconnection is suppressed.  Such
effects can contribute to mode rotation in tokamaks.

Recent PIC simulations by Oka \emph{et al.}\cite{oka:asym}\ displayed
X-line retreat due to the presence of an obstructing wall in one of
the two downstream regions.
The retreat speed was found to be {$\sim$}{$0.1$} of the upstream
Alfv\'en velocity and comparable to the reconnection inflow velocity.
In qualitative agreement with the scaling model developed in Ref.\
\onlinecite{murphy:asym}, the reconnection rate was not significantly
affected by current sheet motion.
However, the structure of the current sheet was modified from the
symmetric case.  In particular, the ion flow stagnation point was
closer to the wall than the X-line, and the outflow jets away from the
obstruction were longer and faster than the outflow jets directed
towards the wall.

In all of these cases, current sheet motion is an important
consideration because it modifies the structure of the reconnection
layer and affects the transport of energy, mass, and momentum.  The
simulations in this paper are used to help understand the impact
current sheet motion has on the structure and dynamics of a
reconnection layer and to determine what sets the rate of X-line
retreat.

This paper is organized as follows.
Section \ref{numerical_method} describes the numerical method used for
the simulations reported in this paper.
Section \ref{problem_setup} describes the initial perturbed
equilibrium and boundary conditions.
Section \ref{case_study} provides a thorough discussion of the
simulation results, including comparisons to a symmetric single
perturbation simulation and details of the internal structure of the
current sheet.
Section \ref{derivation} contains a derivation of the rate of X-line
retreat in terms of local parameters evaluated at the X-point.
Section \ref{conclusions} contains a summary and conclusions.

\section{NUMERICAL METHOD} \label{numerical_method}

The NIMROD code \cite{sovinec:jcp,sovinec:jop,sovinec:2009} (Non-Ideal
Magnetohydrodynamics with Rotation, Open Discussion) is well suited
for the study of resistive MHD and two-fluid magnetic reconnection in
a variety of configurations.\cite{murphy:mrx, hooper:recon,
sovinec:2009} NIMROD uses a finite element formulation for the
poloidal plane and, for three dimensional simulations, a finite
Fourier series in the out-of-plane direction.  
The equations evolved by NIMROD for the two-dimensional simulations
reported in this paper are given in dimensionless form by
\begin{eqnarray}
  \frac{\partial \rho}{\partial t}
  + \nabla \cdot \left( \rho \mathbf{V} \right)
  = \nabla \cdot D \nabla \rho,  \label{continuity}
  \\
  \frac{\partial \mathbf{B}}{\partial t} 
  =
  - \nabla \times 
  \left(
    \eta \mathbf{J} - \mathbf{V}\times\mathbf{B}
  \right), \label{farohms}
  \\
  \mathbf{J} = \nabla \times \mathbf{B}, \label{ampere}
  \\
  \rho 
  \left(
    \frac{\partial \mathbf{V}}{\partial t}
    + \mathbf{V} \cdot \nabla \mathbf{V}
  \right)
  = 
  \mathbf{J}\times\mathbf{B}
  - \nabla{p}
  + \nabla \cdot \rho \nu \nabla \mathbf{V}, \label{momentum}
  \\
  \frac{\rho}{\gamma-1}
  \left(
    \frac{\partial T}{\partial t} + \mathbf{V} \cdot \nabla T
  \right)
  =
  - \frac{p}{2} \nabla \cdot \mathbf{V}
  - \nabla \cdot \mathbf{q}
  + Q, \label{temperature}
\end{eqnarray}
where
$\mathbf{B}$ is the magnetic field, 
$\mathbf{V}$ is the bulk plasma velocity, 
$\mathbf{J}$ is current density, 
$\rho$ is density,
$p$ is the plasma pressure,
$\eta$ is resistivity,
$\nu$ is the kinematic viscosity,
$T$ is temperature,
$\mathbf{q} = -\rho\chi\nabla T$ represents isotropic thermal conduction, 
$\chi$ is the thermal diffusivity, 
$Q$ 
       includes resistive and viscous heating, 
$\gamma=5/3$ is the ratio of specific heats, and
$D$ is an artificial number density diffusivity.
Simulation quantities are normalized to the following respective
values: 
$B_0$, 
$\rho_0$, 
$L_0$,
$t_0$,
$V_{A0} \equiv L_0/t_0 \equiv B_0/\sqrt{\mu_0\rho_0}$,
$p_0 \equiv \rho_0 V_{A0}^2 \equiv B_0/\mu_0 \equiv \rho_0T_0/m_i$, 
$J_0 \equiv B_0/\mu_0L_0$, and
$\eta_0/\mu_0 \equiv \nu_0 \equiv \chi_0 \equiv D_0 \equiv L_0^2/t_0$. 
The divergence constraint is not exactly met, so divergence cleaning
is used to minimize the development of divergence
error.\cite{sovinec:jcp}

NIMROD represents solution fields as the sum of steady-state and
time-varying components.  For most applications the use of an ideal
MHD equilibrium for the steady-state component does not lead to
significant error when a small finite resistivity is used.  However,
in this work it is necessary for the steady-state component to be free
of current for resistive diffusion to be represented accurately.

\section{PROBLEM SETUP} \label{problem_setup}

The initial conditions for the two-dimensional simulations reported in
this paper are those of a perturbed Harris sheet equilibrium.  The
Harris sheet equilibrium is given by
\begin{eqnarray}
  B_x(z) = B_0 \tanh\left(\frac{z}{\delta_0}\right), 
  \label{harris_B} \\
  J_y(z) = \frac{B_0}{\delta_0}
  \sech^2\left(\frac{z}{\delta_0}\right),  
  \label{harris_J}  \\
  p(z) = \frac{B_0^2}{2}
  \left[
    \beta_0 + \sech^2\left(\frac{z}{\delta_0}\right)
  \right]
  \label{harris_p} 
\end{eqnarray}
where $B_0$ is the asymptotic magnetic field strength, 
      $\delta_0$ is the Harris sheet thickness, and 
      $\beta_0$ is the ratio of the asymptotic plasma pressure
      to the asymptotic magnetic pressure.
The initial temperature is constant.
Here, $\hat{\mathbf{x}}$ is the outflow direction, 
      $\hat{\mathbf{y}}$ is the out-of-plane direction,
  and $\hat{\mathbf{z}}$ is the inflow direction.  
The perturbed component of the magnetic field is of the form
\begin{equation}
  \mathbf{B}_p\left(x,z\right) = 
  \nabla \times \left( \psi_p\hat{\mathbf{y}} \right)
\end{equation}
with the flux function $\psi_p$ for double perturbation simulations
given by
\begin{eqnarray}
  \psi_p(x,z) = -B_1 h
  \left\{  
    \exp \left[
           - \left( \frac{x-\Delta}{h} \right)^2
           - \left( \frac{z}{h} \right)^2
         \right] \right. \nonumber \\
  \left.
   +\exp \left[
           - \left( \frac{x+\Delta}{h} \right)^2
           - \left( \frac{z}{h} \right)^2
         \right] 
  \right\},
  \label{perturb_psi}
\end{eqnarray} 
where $B_1$ is the amplitude of the perturbation, $\Delta$ is the
half-separation between the two components of the perturbation, and
$h$ is the length scale of the perturbations.  This configuration is
reminiscent of the final state of the multiple X-line simulations
presented in Ref.\ \onlinecite{nakamura:2010}.  The form of this
perturbation does not depend on the size of the computational domain.

The computational domain is defined by $-x_{max}\leq x \leq x_{max}$ and
$-z_{max}\leq z \leq z_{max}$.  The domain is periodic in the $x$
direction and has no-slip perfectly conducting boundaries at $z=\pm
z_{max}$.  
The domain has $m_x$ and $m_z$ finite elements in the $x$ and $z$
directions, respectively.
Mesh packing is used in both the $x$ and $z$ directions to ensure that
the reconnection layers are sufficiently resolved.
The physical size of the domain is chosen to be large
enough that boundary conditions do not significantly affect the
dynamical behavior over the time scales considered in the analysis.

The simulation parameters are as follows.
The Harris sheet equilibrium is given by 
$B_0 = 1$, 
$\delta_0 = 0.1$, and
$\beta_0 = 1$.  
The initial perturbation is given by
$B_1 = 0.05$, 
$\Delta = 1$, and
$h = 0.5$.  
The length scale is normalized to the half-distance between the
initial perturbations.  
The diffusivities are given by $\eta=\nu=\chi=10^{-3}$ and
$D=10^{-5}$.
All velocities given in this paper are in the simulation reference
frame.
A variable time step is used such that the Courant number is no
greater than $1/6$.
The computational domain is given by
$x_{max} = 30$, 
$z_{max} = 12$, 
$m_x = 124$, and $m_z = 22$   
with sixth order finite element basis functions.
The domain is chosen to be large enough that boundary conditions do
not significantly affect the relevant behavior for the time scales
considered in this paper.
The simulations are ended when the outflow jets reach the
periodic boundary.  Because the simulation is symmetric about $x=0$,
only $x\geq 0$ is considered in the following sections.

\section{SIMULATION RESULTS} \label{case_study}


This section contains a detailed analysis of a double perturbation
simulation which is used to understand the structure and development
of a moving current sheet in the resistive MHD limit.
This double perturbation simulation is compared to an otherwise
equivalent single perturbation simulation which exhibits
non-retreating, symmetric reconnection.  



\subsection{The global structure of the reconnection region}



\begin{figure*} 
  \includegraphics[scale=0.85]{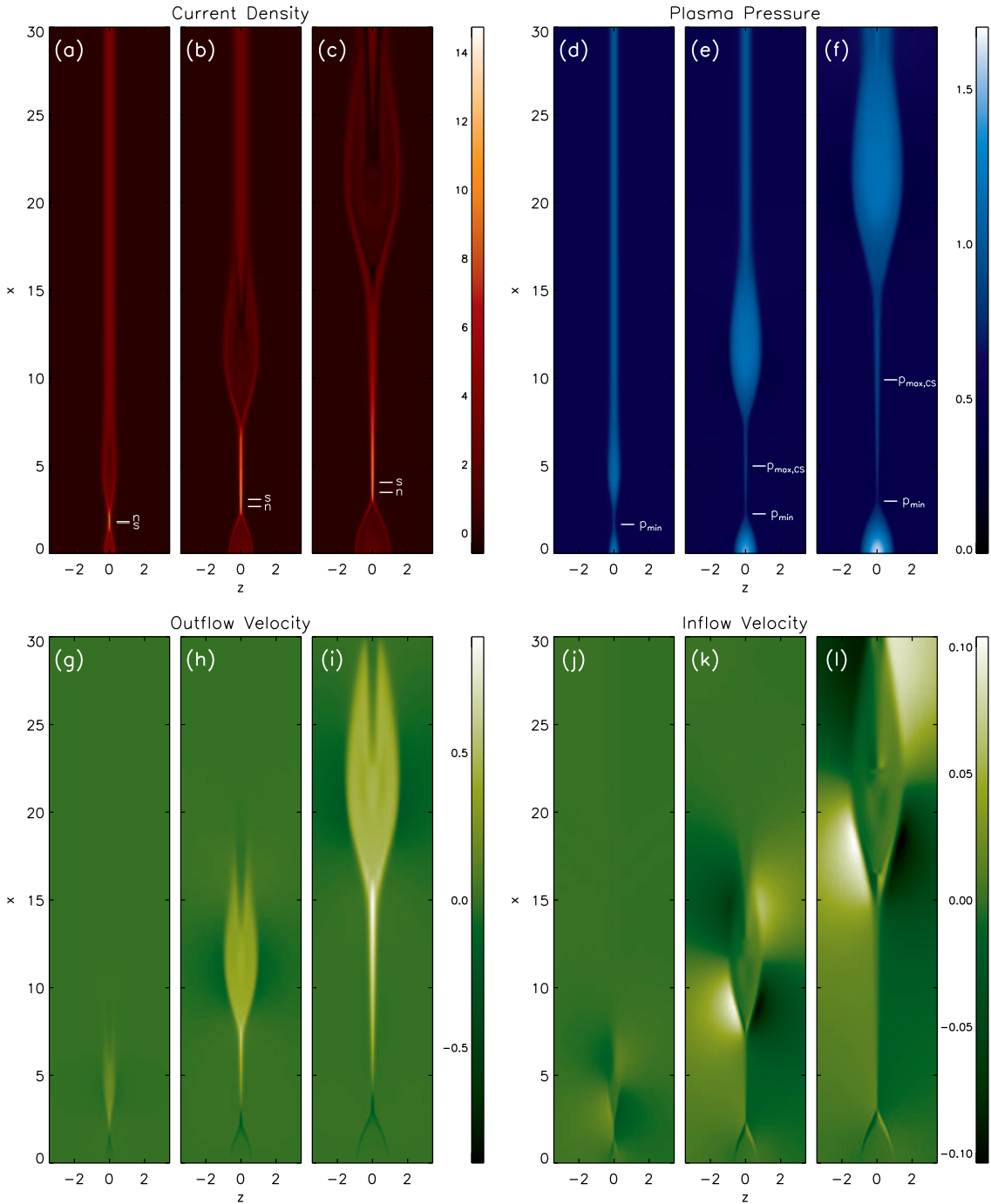}
  \caption{\coline{}Contour plots showing: (a)--(c) the out-of-plane
    current density, $J_y$, (d)--(f) the plasma pressure, $p$,
    (g)--(i) the outflow component of velocity, $V_x$, and (j)--(l)
    the inflow component of velocity, $V_z$, at times $22$, $44$, and
    $66$ for one of the two reconnection regions during the simulation
    of X-line retreat discussed in Sec.\ \ref{case_study}.  The
    locations of the magnetic field null, flow stagnation point,
    pressure minimum along $z=0$, and pressure maximum within the
    current sheet are denoted by `n,' `s,' `$p_{\mathrm{min}}$,' and
    `$p_{\mathrm{max,CS}}$,' respectively.}
  \label{contourfig}
\end{figure*} 

\begin{figure}
  \includegraphics{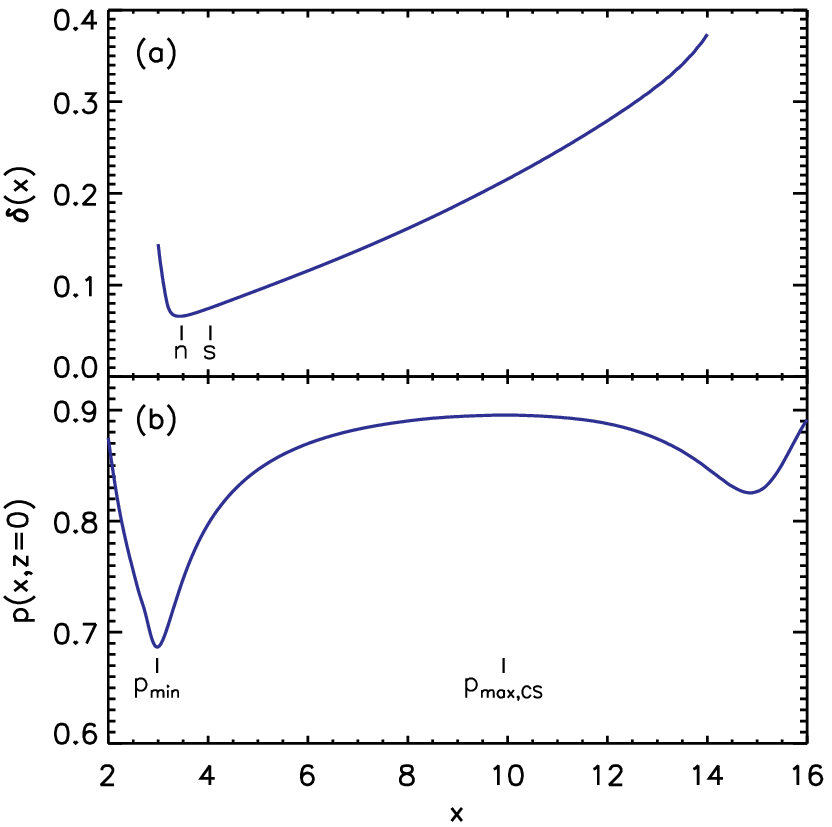}
  \caption{\coline{}Shown above at $t=66$ are (a) the current sheet
  thickness $\delta$ as a function of $x$, calculated as the half
  width at quarter maximum of $J_y$, and (b) the plasma pressure along
  $z=0$.  In (a), the magnetic field null is denoted by `n' and the
  flow stagnation point is denoted by `s,' with $x_n=3.47$ and
  $x_s=4.05$.  In (b), the global pressure minimum along $z=0$ and the
  pressure maximum within the current sheet are denoted by
  `$p_{\mathrm{min}}$' and `$p_{\mathrm{max,CS}}$,' respectively, with 
  $x_{p,\mathrm{min}}=2.98$ and $x_{p,\mathrm{max,CS}} = 9.90$. 
  \label{deltafig}}
\end{figure}

Key features of the time evolution of the double perturbation
simulation are shown in Fig.\ \ref{contourfig}.  The most apparent
feature of the simulations is that the current sheets retreat from
$x=0$ while the reconnection process is developing.  The current sheet
has a single wedge shape which is most apparent late in time.  Figure
\ref{deltafig}(a) shows that the current sheet is thinnest near the
magnetic field null, which is very close to the obstructed exit from
the current sheet, and thickest near the obstructed exit.  


As seen in Figs.\ \ref{contourfig}(d)--(f), the peak plasma pressure
in the magnetic island around $x=0$ is greater than the peak plasma
pressure in the downstream region away from $x=0$.  The plasma
pressure gradient associated with the obstructing magnetic island is
stronger and more localized than the plasma pressure gradient
associated with the unobstructed outflow region.  
Early in time ($t\lesssim 20$), the X-point is located near the
minimum of plasma pressure along $z=0$.  Later ($20 \lesssim t
\lesssim 28$), the plasma pressure at the flow stagnation point is
less than at the X-point.
A local maximum of plasma pressure appears at $t\approx 27$ in the
central region of the current sheet which persists through the
remainder of the simulation.  This pressure maximum is located
slightly closer to the unobstructed exit of the current sheet than the
obstructed exit.

%

Contours of the outflow component of velocity presented in Figs.\
\ref{contourfig}(g)--(i) show that the outflow jet directed away from
the obstruction is faster and longer than the outflow jet directed
towards the obstruction.  This behavior is consistent with previous
simulations of asymmetric outflow reconnection.\cite{roussev:2001,
galsgaard:2002, oka:asym, zhu:2009, nakamura:2010, reeves:2010} 
The kinetic energy and enthalpy fluxes away from the obstruction are
much greater than the those towards the obstruction.

The inflow component of velocity is shown in Figs.\
\ref{contourfig}(j)--(l).  In the upstream regions outside the current
sheet, the inflow component of velocity is roughly uniform except near
the outflow region furthest from $x=0$.  The inflow velocity late in time is
$\sim${$0.015$}--{$0.02$}.  Because the large outflow blob from the
unobstructed exit of the current sheet seen in Fig.\ \ref{contourfig}
is propagating away from $x=0$, there is a quadrupole-like pattern for
$V_z$ as upstream plasma ahead of the blob moves out of the way and
upstream plasma behind the blob is pulled back towards $z=0$.

\begin{figure}
  \includegraphics[scale=0.95]{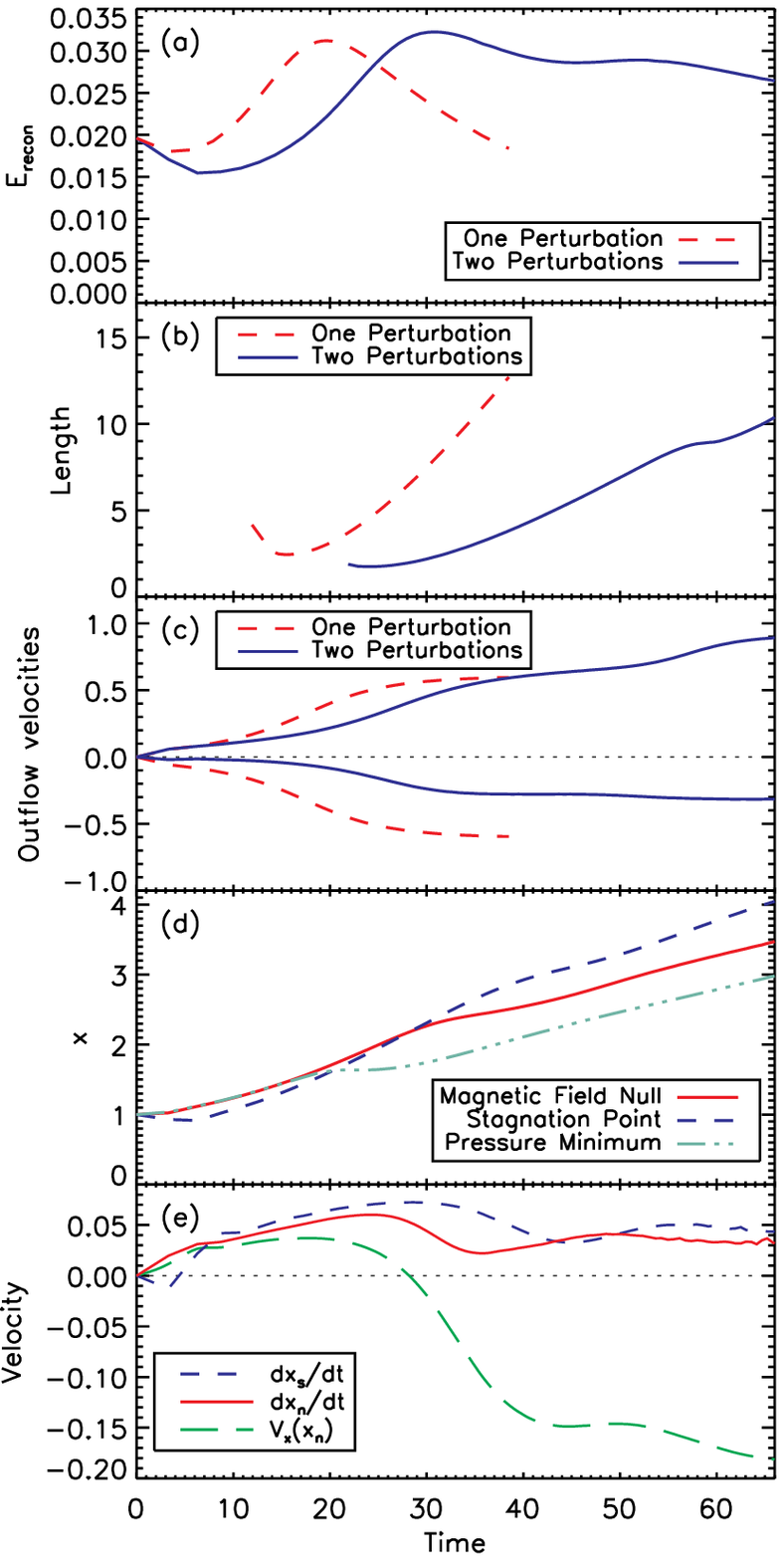}
  \caption{\coline{}Simulation results showing as a function of time
    (a) the reconnection electric field strength measured at the
    X-point, (b) the full length at quarter maximum of $J_y$ along
    $z=0$, (c) the peak outflow velocities, (d) the positions of the
    magnetic field null, flow stagnation point, and pressure minimum
    along $z=0$ ($x_n$, $x_s$, and $x_{p,\mathrm{min}}$,
    respectively), and (e) the rates of change in position of the
    magnetic field null and flow stagnation point ($\dif x_n/\dif t$
    and $\dif x_s/\dif t$, respectively) and the velocity at the
    magnetic field null, $V_x(x_n)$.  }
  \label{timefig}
\end{figure}

Figure \ref{timefig} shows the time evolution of key quantities from
the double perturbation simulation with comparisons to the otherwise
equivalent single perturbation simulation used as a control for this
study.  Because a magnetic island forms at $t\approx 38$ in the single
perturbation simulation, the comparison is halted at that time.
Figure \ref{timefig}(a) shows the reconnection electric field strength
for both simulations.
The reconnection rate peaks more quickly in the single perturbation
simulation than the double perturbation simulation.  The development
of reconnection in the double perturbation simulation is slow because
the magnetic field null is located near the global pressure minimum
along $z=0$ for $t\lesssim 20$ [see Fig.\ \ref{timefig}(d)] so that
pressure gradients oppose outflow instead of facilitating it.
However, the amplitude of the reconnection rate is eventually greater
in the double perturbation case because the current sheet can increase
in length in only one direction.  As seen in Fig.\ \ref{timefig}(b),
the single perturbation current sheet is a factor of
$\sim${$2$}--{$3$} longer at any given time.  The slight disturbances
around $t\gtrsim 58$ in Figs.\ \ref{timefig}(a) and \ref{timefig}(b)
may be due to the periodic boundary condition.  While the reconnection
rate is enhanced slightly in the double perturbation case, the
difference is less than a factor of two so we conclude that the
reconnection rate is only modestly affected by asymmetry or current
sheet motion (see also Refs.\ \onlinecite{murphy:asym} and
\onlinecite{oka:asym}).

The peak outflow velocities in the positive and negative directions
are shown in Fig.\ \ref{timefig}(c).  Again, it is apparent that the
double perturbation simulation takes longer to develop than the single
perturbation simulation.  The unobstructed outflow jet is typically
$\sim${$2$}--{$3$} times faster than the obstructed outflow jet.  

\subsection{The internal structure of the current sheet}

\begin{figure}
  \includegraphics{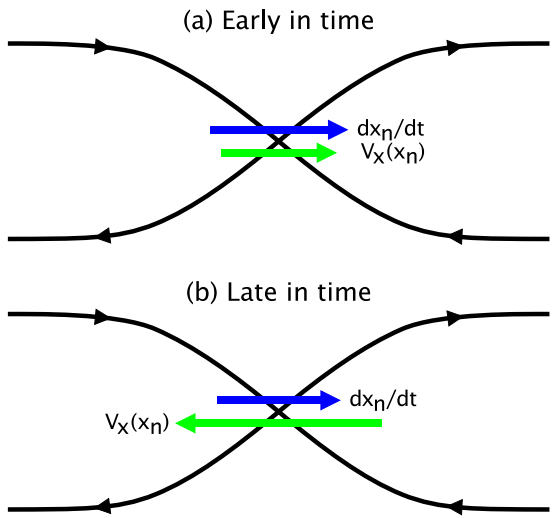}
  \caption{\coline{}A schematic showing the rate of X-line retreat,
    $\dif x_n/\dif t$, and the plasma flow velocity at the X-line,
    $V_x(x_n)$.  Early in time, both velocities are in the same
    direction.  Late in time, there is significant plasma flow across
    the X-line in the opposite direction of X-line retreat.
    \label{directions}}
\end{figure}

The positions of the magnetic field null, flow stagnation point, and
pressure minimum along $z=0$ ($x_n$, $x_s$, and $x_{p,\mathrm{min}}$,
respectively) are presented in Fig.\ \ref{timefig}(d).  The positions
of the X-point and flow stagnation point are separated by a short
distance.  Both points are located very close to the obstructed exit
of the current sheet.  Early in time, the flow stagnation point is
closer to the obstruction than the magnetic field null (e.g.,
$x_s<x_n$).  This corresponds to the plasma flow at the X-line moving
in the same direction as X-line retreat.  Such a scenario is what one
would expect if the frozen-in condition is approximately met [see
Fig.\ \ref{directions}(a)].  However, around $t\approx 28$ the
relative positions of the magnetic field null and flow stagnation
point switch (e.g., $x_s>x_n$).  Hence at late times during this
simulation the plasma flow at the X-line is in the opposite direction
of X-line retreat [see Fig.\ \ref{directions}(b)].  The peak resistive
electric field in the double perturbation simulation occurs slightly
after the time when $x_n=x_s$ (e.g., near the time when tension forces
entirely work towards accelerating outflow rather than partially
working against it).

The relative positions of the X-line and flow stagnation point switch
because during near steady conditions, the flow stagnation point will
in general be located near where the tension and pressure gradient
forces cancel (e.g., Ref.\ \onlinecite{murphy:asym}).  This implies
that the total pressure at the flow stagnation point should be greater
than the total pressure at the X-point when time-dependent effects are
not important.  Early in time ($t \lesssim 20$), the X-point is
located very near the global pressure minimum along $z=0$ [see Fig.\
\ref{timefig}(d)].  During this phase, $p(x_s)>p(x_n)$ and the
pressure gradient and tension forces at the flow stagnation point are
oppositely directed.  As the simulation progresses, the X-line and
flow stagnation point retreat more quickly than the position of this
pressure minimum.  For $20\lesssim t\lesssim 28$, $p(x_s)<p(x_n)$ and
the tension and pressure gradient forces at the flow stagnation point
are pointed in the same direction and thus cannot cancel each other
out.  The flow stagnation point retreats quickly until $x_n<x_s$ and
$p(x_s)>p(x_n)$ once again and for the remainder of the simulation
[see Fig.\ \ref{deltafig}(b)].


Figure \ref{timefig}(e) compares the rate of change in position of the
magnetic field null $\dif x_n / \dif t$, the rate of change in
position of the flow stagnation point $\dif x_s / \dif t$, and the
plasma flow velocity across the magnetic field null $V_x(x_n)$.  As
discussed in Ref.\ \onlinecite{seaton:2008} and Sec.\
\ref{derivation}, any difference between $V_x(x_n)$ and $\dif x_n /
\dif t$ must be due to resistive diffusion.  The X-line is observed to
retreat at a variable velocity ranging between $0.02$ and $0.06$.  The
flow stagnation point retreats more quickly than the magnetic field
null for most of the simulation with velocities ranging from $0.03$ to
$0.07$.  For much of the first third of the simulation, $\dif x_s /
\dif t \sim \dif x_n / \dif t \sim V_x(x_n)$.  When $x_s \approx x_n$,
the rate of X-line retreat drops by a factor of $\sim${$3$} before
slowly increasing and stabilizing.  Following a delay after the drop
in $\dif x_n/\dif t$, $\dif x_s / \dif t$ declines also before
stabilizing.  However, the most striking feature of Fig.\
\ref{timefig}(e) is that late in time there is significant plasma flow
across the X-line in the opposite direction of X-line retreat:
$V_x(x_n)\sim -0.16$ while $\dif x_n/\dif t \sim 0.04$.  

\begin{figure}
  \includegraphics{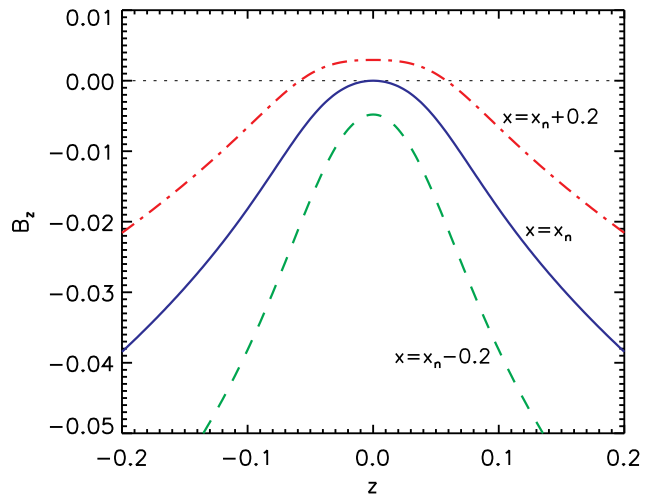}
  \caption{\coline{}The normal component of the magnetic field as a
    function of $z$ for three locations surrounding the magnetic field
    null at $t=66$. At $x=x_n$, $B_z<0$ in the vicinity of the current
    sheet except at $z=0$.
  \label{bznull}}
\end{figure}

\begin{figure}
  \includegraphics{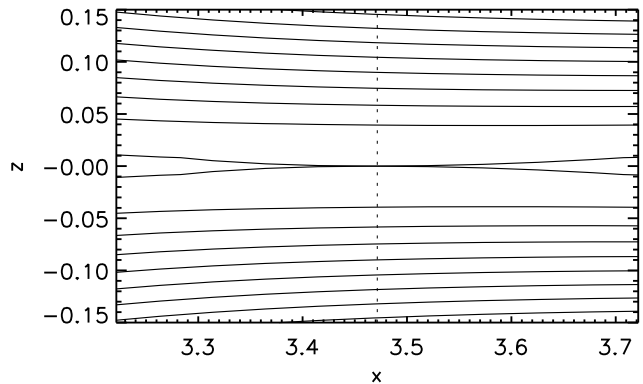}
  \caption{\coline{}Magnetic flux contours near the magnetic field
    null at $t=66$.  The dotted line represents $x=x_n$.
  \label{flux}}
\end{figure}

Figure \ref{bznull} shows $B_z(z)$ for three locations surrounding the
magnetic field null.  Along $x=x_n$, $B_z<0$ except at $z=0$.
Additionally, $\partial^2 B_z/\partial t^2<0$ in the vicinity of the
X-point.  This magnetic field structure near the X-point can be seen
clearly in the magnetic flux contour plot presented in Fig.\
\ref{flux}, where magnetic field lines are pinched inward towards the
right of the X-point.  As we shall see in Sec.\ \ref{derivation},
these features facilitate X-line retreat by diffusion of the normal
component of the magnetic field in the inflow direction.

Plasma flow at the X-line in the direction opposite to X-line retreat
appears to be a persistent feature for the assumed configuration.
This effect occurs during analogous simulations with initial uniform
guide fields up to at least $B_{y0}=4$, a line-tied boundary at $x=0$,
$\beta_0$ at least between $0.25$ and $4$, different diffusivities,
and different initial separations at least for $\Delta<8$.  During
some of these simulations, an additional X-line appears near the
leading end of the current sheet.
For larger $\Delta$, the developing current sheets may become unstable
to the plasmoid instability\cite{loureiro:2007, samtaney:2009,
bhattacharjee:2009, huang:2010, ni:2010, shepherd:2010} before the
retreat process begins.
Plasma flow across the X-line in the direction opposite to
X-line retreat does occasionally occur during analogous resistive MHD
simulations of multiple competing reconnection sites, but less
frequently because current sheet motion is
inhibited.\cite{young:pc:2010}

\subsection{Components of the electric field and momentum balance}

\begin{figure}
  \includegraphics{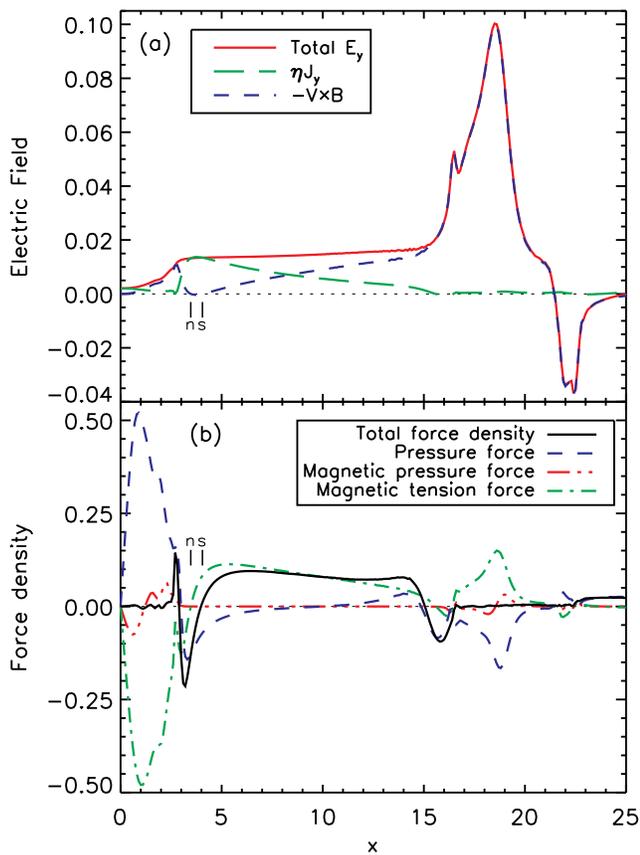}
  \caption{\coline{}Simulation results showing (a) components of the
    electric field and (b) force density contributions along $z=0$ at
    $t=66$.  The magnetic field null is denoted by `n' and the flow
    stagnation point is denoted by `s.' 
  }
  \label{electric_mom}
\end{figure}


Components of the out-of-plane electric field along $z=0$ are shown in
Fig.\ \ref{electric_mom}(a) for $t=66$.  This is late in the
simulation when reconnection is well-developed and $x_n<x_s$.  Inside
the current sheet ($3\lesssim x \lesssim 14$), the electric field is
approximately uniform but increases slightly with distance from $x=0$.
The resistive electric field peaks in between the magnetic field null
and flow stagnation point.  In the obstructed outflow region, the
electric field drops to a fraction of the value from inside the
current sheet.  From Faraday's law, the positive slope of $E_y$
indicates that $B_z$ in the obstructing magnetic island is increasing
with time.  In the unobstructed outflow region, $E_y$ peaks
sharply. This signature is indicative of the unobstructed downstream
plasmoid being advected away from $x=0$.

Figure \ref{electric_mom}(b) shows force density contributions from
the momentum equation (Eq.\ \ref{momentum}) along $z=0$ at $t=66$.
Contributions include the magnetic tension force $B_z\partial
B_x/\partial z$, the magnetic pressure gradient force $-\partial
(B_z^2/2)/\partial x$, and the plasma pressure gradient force
$-\partial p/\partial x$.  Viscous forces are small and not shown
explicitly, but are included in the calculation for total force
density.  The outflow towards the unobstructed exit is accelerated
primarily by magnetic tension, whereas the outflow towards $x=0$ is
accelerated primarily by the pressure gradient force.  Intuitively,
this is because the X-line is located very near one end of the current
sheet so that the tension force directed towards that end is small and
the tension force directed away from that end is large (e.g., Ref.\
\onlinecite{murphy:mrx}).  The force density at the flow stagnation
point is negative until $t\approx 39$ when it becomes positive for the
remainder of the simulation.

\section{THE RATE OF X-LINE RETREAT\label{derivation}}


This section contains a derivation of the rate of X-line retreat in
the two-dimensional case.  The geometry is assumed to be the same as
that of the simulations reported in this paper.  In particular,
symmetry is assumed about $z=0$ so that the motion of the X-line is
purely in the $x$ direction.  No assumptions are made about the
presence or absence of a guide field, except that $\partial/\partial y
\rightarrow 0$.  The analysis presented in this
section can also be applied to the asymmetric inflow case by switching
the inflow and outflow directions (see also Ref.\
\onlinecite{cassak:dissipation}).  


The derivation of the rate of retreat of the X-line begins with
Faraday's law,
\begin{equation}
  \frac{\partial \mathbf{B}}{\partial t} 
  = - \nabla\times\mathbf{E}.
\end{equation}
Evaluating the $z$-component of this expression and using the
assumption of symmetry in the out-of-plane direction gives
\begin{equation}
  \frac{\partial B_z}{\partial t}
  = 
  - \frac{\partial E_y}{\partial x}.\label{faradayz}
\end{equation}
Next, the convective derivative of $B_z$ at the X-point taken using
the velocity of X-line retreat $\dif x_n/\dif t$ is given by
\begin{equation}
  \left.\frac{\partial B_z}{\partial t}\right|_{x_n} +
  \frac{\dif x_n}{\dif t}
  \left.\frac{\partial B_z}{\partial x}\right|_{x_n}
  = 0.\label{geomn}
\end{equation}
The right hand side of Eq.\ \ref{geomn} is zero because the normal
component of the magnetic field at the X-point does not change from
zero, by definition.  The rate of X-line retreat then follows from
Eqs.\ \ref{faradayz} and \ref{geomn} and is given by
\begin{equation}
  \frac{\dif x_n}{\dif t} = 
  \left.
  \frac{\partial E_y/\partial x}{\partial B_z/\partial x}
  \right|_{x_n}.\label{dxndt1}
\end{equation}
Intuitively, this means that the X-line moves in the direction of
increasing reconnection electric field strength.  This expression
shows that it is not self-consistent to assume that the X-line is
moving while the out-of-plane electric field is constant in space
(cf.\ Ref.\ \onlinecite{owen:1987:451,*owen:1987:467}).

This derivation has thus far not specified an expression for the
out-of-plane electric field.  By using the resistive MHD Ohm's law
$\mathbf{E} + \mathbf{V}\times\mathbf{B}=\eta\mathbf{J}$, Eq.\
\ref{dxndt1} yields
\begin{equation}
  \frac{\dif x_n}{\dif t}
  = V_x(x_n) - \eta 
  \left[
    \frac{\frac{\partial^2 B_z}{\partial x^2} +
          \frac{\partial^2 B_z}{\partial z^2}
         }{\frac{\partial B_z}{\partial x}         
         }
  \right]_{x_n}. \label{dxndt}
\end{equation}
For the assumed geometry, this is an exact result for resistive MHD\@.
This analysis can be extended to include additional terms in the
generalized Ohm's law.  The inverse dependence on $\left. \partial
B_z/\partial x \right|_{x_n}$ in the resistive term comes from the
geometric properties of Eq.\ \ref{geomn}; it is easier to change the
position of a root of a function by a vertical shift if the slope of
the function near the root is shallow.  Because $\left. \partial^2 B_z
/ \partial x^2 \right|_{x_n}$ will usually be small if
$\left. \partial B_z / \partial x \right|_{x_n}$ is small, it is
likely that in resistive MHD the term $\left. \partial^2 B_z/\partial
z^2 \right|_{x_n}$ representing diffusion of the normal component of
the magnetic field in the inflow direction will be more important than
the term $\left. \partial^2 B_z/\partial x^2 \right|_{x_n}$
representing diffusion of the normal component of the magnetic field
along the outflow direction.  Indeed, the simulation does show that
late in time $\left. \partial^2 B_z/\partial x^2 \right|_{x_n} \ll
\left. \partial^2 B_z/\partial z^2 \right|_{x_n}$.  When the X-line is
retreating during collisionless reconnection, it is possible that the
term $\partial^2 B_z/\partial x^2$ will become more important.  During
three-dimensional reconnection, the term $\partial^2 B_z/\partial y^2$
may also play a role in facilitating X-line retreat.


\begin{figure}
  \includegraphics[scale=1]{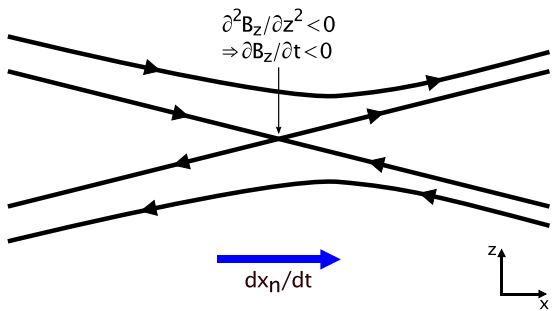}
  \caption{\coline{}The mechanism for X-line retreat due to diffusion
    of the normal component of the magnetic field in the inflow
    direction.
  \label{mechanism}}
\end{figure}

The mechanism for diffusion of the X-line is presented in Fig.\
\ref{mechanism}.  Along $x=x_n$, $B_z<0$ except at $z=0$.  Then
because $\partial^2 B_z/\partial z^2 < 0$, the normal component of the
magnetic field becomes more negative in the immediate vicinity of the
X-point.  This diffusion of $B_z$ in the $z$ direction causes the
magnitude of $B_z$ to become stronger immediately to the left of the
X-point and weaker immediately to the right of the X-point in Fig.\
\ref{mechanism}.  As a consequence, the X-point retreats to the right.
The necessary features for this diffusion mechanism are apparent in
Figs.\ \ref{bznull} and \ref{flux}.  The contribution to X-line motion
from diffusion of the normal component of the magnetic field in the
outflow direction is not shown in Fig.\ \ref{mechanism}.

The difference between the bulk plasma flow across the X-line,
$V_x(x_n)$, and the rate in change of position of the X-line, $\dif
x_n / \dif t$, is subtle but important.  In the ideal MHD limit, the
magnetic field will be purely frozen in to the plasma so that these
two quantities will be identical [e.g., $V_x(x_n) = \dif x_n / \dif
t$].  In this limit the X-line is purely advected by the bulk plasma
flow.  Therefore, any difference between $V_x(x_n)$ and $\dif x_n/\dif
t$ must be due to resistive diffusion of the magnetic field (see also
Ref.\ \onlinecite{seaton:2008}).

The relation presented in Eq.\ \ref{dxndt} and Fig.\ \ref{mechanism}
shows that the rate of X-line retreat depends fundamentally on local
parameters near the X-point.  Hence global models
attempting to describe the rate of X-line retreat must take into
account the coupling between large scales and the local structure of
the magnetic field near the X-point.  A starting point for such models
may be the analysis presented in Ref.\ \onlinecite{shibata:2001}.

\section{DISCUSSION AND CONCLUSIONS\label{conclusions}}

This paper presents a resistive MHD simulation of two competing
reconnection sites which move apart from each other as they develop.
This investigation provides insight into the impact of current sheet
motion on the reconnection process as well as on what happens when
outflow from one exit of the current sheet is blocked.

The asymmetry in the reconnection process is most apparent in the
outflow velocity profile.  When the reconnection process is
well-developed, the unobstructed outflow jet is {$\sim$}$2$--$3$ times
faster than the obstructed outflow jet.  As a consequence, most of the
mass, energy, and momentum flux associated with the outflow is
directed away from the obstructing magnetic island that forms between
the two current sheets.  Late in time, the reconnection rate is
slightly higher in the double perturbation simulation than in an
otherwise equivalent single perturbation simulation because the
obstruction prevents the length of each current sheet from growing in
one direction.

The X-point and flow stagnation point are both located very near the
obstructed exit of the current sheet.  This gives the reconnection
layer a characteristic single wedge shape which is especially apparent
late in time.  
During recent simulations of the plasmoid
instability,\cite{loureiro:2007, samtaney:2009, bhattacharjee:2009,
huang:2010, ni:2010, shepherd:2010} single wedge shaped small-scale
reconnection sites are routinely observed with the X-point located
near the thinnest part of the current sheet (see the online movie
associated with Fig.\ 3 of Ref.\ \onlinecite{huang:2010}).
Because the X-line is located very close to one end of the current
sheet, the tension force is directed predominantly towards the other
direction (see also Refs.\ \onlinecite{murphy:mrx} and
\onlinecite{galsgaard:2002}).  As in previous simulations of
reconnection with asymmetric inflow\cite{cassak:asym, cassak:hall,
cassak:dissipation, labellehamer:1995, ugai:2000, kondoh:2004,
borovsky:asym, birn:2008, pritchett:2008, birn:2010, murphy:mrx} and
asymmetric outflow,\cite{oka:asym, murphy:mrx, murphy:asym} the X-line
and flow stagnation point are separated by a short distance.  This
separation appears to be a ubiquitous feature of asymmetric
reconnection.

The X-line retreats at speeds of $\sim${$0.02$}--{$0.06$} and the flow
stagnation point retreats at speeds of $\sim${$0.03$}--{$0.07$}.  This
is slower than the X-line retreat speed of {$\sim$}$0.1$ observed in
Ref.\ \onlinecite{oka:asym}.  Early in time, the plasma flow at the
X-line is in the same direction as the rate in change of position of
the X-line.  While the reconnection process is still developing,
however, the relative positions of the X-line and flow stagnation
point switch.  As a consequence, late in time the plasma flow at the
X-line is in the opposite direction of X-line retreat.  This switch
occurs so that the flow stagnation point will be located near where the
magnetic tension and plasma pressure forces cancel.

To further understand these results, an expression is derived for the
rate of X-line retreat.  In the assumed geometry, the X-line retreats
in the direction of increasing reconnection electric field strength.
In the resistive MHD limit, the X-line retreats due to either
advection by the bulk plasma flow or by diffusion of the normal
component of the magnetic field.  

Interestingly, previous PIC simulations show that while the ion flow
at the X-point is in the same direction as X-line retreat, the
electron flow at the X-point is in the opposite direction (see Fig.\ 1
of Ref.\ \onlinecite{oka:asym}).  Because in Hall MHD the magnetic
field is frozen into the electron fluid rather than the bulk plasma,
this result suggests that a similar effect occurs in fully kinetic
simulations.  It will be important in future work to determine whether
or not this class of behavior occurs in more realistic geometries such
as CME current sheets and the Earth's magnetotail, or for substantially
different plasma parameters.

The simulation presented in this paper can be used to assess the
validity of the assumptions made by the steady-state model of
reconnection with asymmetry in the outflow direction presented in
Ref.\ \onlinecite{murphy:asym}.  In particular, two assumptions
require refinement.  The first assumption is that the magnetic tension
force contributes evenly to outflow from both sides.  In the
simulation, however, the tension force is almost entirely directed
towards the unobstructed downstream region because the X-line is
located very close to the obstructed exit.  The second assumption is
that the current sheet has approximately uniform thickness along the
outflow direction.  However, Figs.\ \ref{contourfig}(c) and
\ref{deltafig}(a) show that the current sheet has a characteristic single
wedge shape.  Consequently, models for asymmetric outflow reconnection
will need to be refined to account for these effects.




The results of this paper may have several implications for current
sheets that form in the wakes of CMEs.  In these events, the locations
of the predominant X-line and flow stagnation point are expected to be
located near the base of the current sheet (see also Refs.\
\onlinecite{savage:2010}, \onlinecite{reeves:2010}, and
\onlinecite{seaton:2008}).  Consequently, the mass, momentum, and
energy fluxes are expected to be greater in the antisunward direction.
This may provide an explanation for some of the observed properties of
CMEs: that the masses of CMEs increase after leaving the
Sun,\cite{bemporad:2007, vourlidas:2010, lin:2004} and that CMEs are
heated even after leaving the flare site.\cite{akmal:2001,
ciaravella:2001, lee:2009, landi:2010} Additionally, current sheets
behind CMEs are expected to be thinner when observed at low altitudes
and thicker when observed at high altitudes (e.g., compare the
inferred thicknesses in Refs.\ \onlinecite{2008ApJ...686.1372C} and
\onlinecite{savage:2010}).  However, more observational and
theoretical work is necessary before drawing conclusions.  A
complicating factor is that CME current sheets may be susceptible to
the tearing\cite{fkr, 2007ApJ...658L.123L} and
plasmoid\cite{loureiro:2007, samtaney:2009, bhattacharjee:2009,
huang:2010, ni:2010, shepherd:2010} instabilities.

Effects associated with competition between multiple competing
reconnection sites are likely to have consequences on the physics of
turbulent reconnection. The model for turbulent reconnection by
Lazarian and Vishniac\cite{lazarian:turbrecon} assumes that outflows
from each of the small-scale reconnection sites in a large-scale
current sheet do not significantly affect other reconnection sites.
The two-dimensional simulations reported in this paper cannot fully
address this problem.  Hence in future work we will perform
three-dimensional simulations with the initial perturbations offset
from each other in the out-of-plane direction.

The resistive MHD simulations presented in this paper are not directly
applicable to magnetic reconnection in the Earth's magnetotail.
There, two-fluid and collisionless effects not included in the
resistive MHD framework are known to be important.  Fully kinetic
simulations such as those presented in Ref.\ \onlinecite{oka:asym}
show many of the same features as the double perturbation simulation
presented here.  
Moreover, the shape and structure of the out-of-plane quadrupole
magnetic field associated with two-fluid reconnection can be greatly
affected by X-line retreat.\cite{oka:asym, oka:pc:2010} Hence, the
NIMROD simulations reported in this paper will be extended to include
two-fluid effects.



\begin{acknowledgments}

The author thanks J.~C.\ Raymond, C.~R.\ Sovinec, E.~G.\ Zweibel,
P.~A.\ Cassak, J.\ Lin, C.~C.\ Shen, M.\ Oka, D.\ Seaton, K.~K.\
Reeves, A.~K.\ Young, B.~P.\ Sullivan, J.~M.\ Stone, and S.\ Friedman
for useful discussions; Y.-M.\ Huang for an intuitive understanding of
Eq.\ \ref{geomn}; and the members of the NIMROD Team for code
development efforts that made this work possible.  N.~A.~M.\
acknowledges the hospitality of the Yunnan Astronomical Observatory
during a visit when much of this paper was written.  This research has
benefited from the use of NASA's Astrophysics Data System.
  
This research was supported by NASA grant NNX09AB17G to the
Smithsonian Astrophysical Observatory and a grant from the Smithsonian
Institution Sprague Endowment Fund during FY10.  N.~A.~M.\
also acknowledges support from the Center for Magnetic
Self-Organization in Laboratory and Astrophysical Plasmas while a
graduate student at the University of Wisconsin when this project was
initiated.

\end{acknowledgments}


\begin{thebibliography}{67}%
\makeatletter
\providecommand \@ifxundefined [1]{%
 \@ifx{#1\undefined}
}%
\providecommand \@ifnum [1]{%
 \ifnum #1\expandafter \@firstoftwo
 \else \expandafter \@secondoftwo
 \fi
}%
\providecommand \@ifx [1]{%
 \ifx #1\expandafter \@firstoftwo
 \else \expandafter \@secondoftwo
 \fi
}%
\providecommand \natexlab [1]{#1}%
\providecommand \enquote  [1]{``#1''}%
\providecommand \bibnamefont  [1]{#1}%
\providecommand \bibfnamefont [1]{#1}%
\providecommand \citenamefont [1]{#1}%
\providecommand \href@noop [0]{\@secondoftwo}%
\providecommand \href [0]{\begingroup \@sanitize@url \@href}%
\providecommand \@href[1]{\@@startlink{#1}\@@href}%
\providecommand \@@href[1]{\endgroup#1\@@endlink}%
\providecommand \@sanitize@url [0]{\catcode `\\12\catcode `\$12\catcode
  `\&12\catcode `\#12\catcode `\^12\catcode `\_12\catcode `\%12\relax}%
\providecommand \@@startlink[1]{}%
\providecommand \@@endlink[0]{}%
\providecommand \url  [0]{\begingroup\@sanitize@url \@url }%
\providecommand \@url [1]{\endgroup\@href {#1}{\urlprefix }}%
\providecommand \urlprefix  [0]{URL }%
\providecommand \Eprint [0]{\href }%
\@ifxundefined \urlstyle {%
  \providecommand \doi  [0]{\begingroup \@sanitize@url \@doi}%
  \providecommand \@doi [1]{\endgroup \@@startlink {\doibase
  #1}doi:\discretionary {}{}{}#1\@@endlink }%
}{%
  \providecommand \doi  [0]{doi:\discretionary{}{}{}\begingroup
  \urlstyle{rm}\Url }%
}%
\providecommand \doibase [0]{http://dx.doi.org/}%
\providecommand \Doi [0]{\begingroup \@sanitize@url \@Doi }%
\providecommand \@Doi  [1]{\endgroup\@@startlink{\doibase#1}\@@Doi}%
\providecommand \@@Doi [1]{#1\@@endlink}%
\providecommand \selectlanguage [0]{\@gobble}%
\providecommand \bibinfo  [0]{\@secondoftwo}%
\providecommand \bibfield  [0]{\@secondoftwo}%
\providecommand \translation [1]{[#1]}%
\providecommand \BibitemOpen [0]{}%
\providecommand \bibitemStop [0]{}%
\providecommand \bibitemNoStop [0]{.\EOS\space}%
\providecommand \EOS [0]{\spacefactor3000\relax}%
\providecommand \BibitemShut  [1]{\csname bibitem#1\endcsname}%
\bibitem [{\citenamefont {Murphy}\ \emph {et~al.}(2010)\citenamefont {Murphy},
  \citenamefont {Sovinec},\ and\ \citenamefont {Cassak}}]{murphy:asym}%
  \BibitemOpen
  \bibfield  {author} {\bibinfo {author} {\bibfnamefont {N.~A.}\ \bibnamefont
  {Murphy}}, \bibinfo {author} {\bibfnamefont {C.~R.}\ \bibnamefont {Sovinec}},
  \ and\ \bibinfo {author} {\bibfnamefont {P.~A.}\ \bibnamefont {Cassak}},\
  }\href@noop {} {\bibfield  {journal} {\bibinfo  {journal} {J. Geophys.
  Res.},\ }\textbf {\bibinfo {volume} {115}},\ \bibinfo {eid} {A09206}
  (\bibinfo {year} {2010})}\BibitemShut {NoStop}%
\bibitem [{\citenamefont {{Baker}}\ \emph {et~al.}(1996)\citenamefont
  {{Baker}}, \citenamefont {{Pulkkinen}}, \citenamefont {{Angelopoulos}},
  \citenamefont {{Baumjohann}},\ and\ \citenamefont
  {{McPherron}}}]{baker:1996}%
  \BibitemOpen
  \bibfield  {author} {\bibinfo {author} {\bibfnamefont {D.~N.}\ \bibnamefont
  {{Baker}}}, \bibinfo {author} {\bibfnamefont {T.~I.}\ \bibnamefont
  {{Pulkkinen}}}, \bibinfo {author} {\bibfnamefont {V.}~\bibnamefont
  {{Angelopoulos}}}, \bibinfo {author} {\bibfnamefont {W.}~\bibnamefont
  {{Baumjohann}}}, \ and\ \bibinfo {author} {\bibfnamefont {R.~L.}\
  \bibnamefont {{McPherron}}},\ }\href@noop {} {\bibfield  {journal} {\bibinfo
  {journal} {J. Geophys. Res.},\ }\textbf {\bibinfo {volume} {101}},\ \bibinfo
  {pages} {12975} (\bibinfo {year} {1996})}\BibitemShut {NoStop}%
\bibitem [{\citenamefont {{Runov}}\ \emph {et~al.}(2003)\citenamefont
  {{Runov}}, \citenamefont {{Nakamura}}, \citenamefont {{Baumjohann}},
  \citenamefont {{Treumann}}, \citenamefont {{Zhang}}, \citenamefont
  {{Volwerk}}, \citenamefont {{V{\"o}r{\"o}s}}, \citenamefont {{Balogh}},
  \citenamefont {{Gla{\ss}meier}}, \citenamefont {{Klecker}}, \citenamefont
  {{R{\`e}me}},\ and\ \citenamefont {{Kistler}}}]{runov:2003}%
  \BibitemOpen
  \bibfield  {author} {\bibinfo {author} {\bibfnamefont {A.}~\bibnamefont
  {{Runov}}}, \bibinfo {author} {\bibfnamefont {R.}~\bibnamefont {{Nakamura}}},
  \bibinfo {author} {\bibfnamefont {W.}~\bibnamefont {{Baumjohann}}}, \bibinfo
  {author} {\bibfnamefont {R.~A.}\ \bibnamefont {{Treumann}}}, \bibinfo
  {author} {\bibfnamefont {T.~L.}\ \bibnamefont {{Zhang}}}, \bibinfo {author}
  {\bibfnamefont {M.}~\bibnamefont {{Volwerk}}}, \bibinfo {author}
  {\bibfnamefont {Z.}~\bibnamefont {{V{\"o}r{\"o}s}}}, \bibinfo {author}
  {\bibfnamefont {A.}~\bibnamefont {{Balogh}}}, \bibinfo {author}
  {\bibfnamefont {K.}~\bibnamefont {{Gla{\ss}meier}}}, \bibinfo {author}
  {\bibfnamefont {B.}~\bibnamefont {{Klecker}}}, \bibinfo {author}
  {\bibfnamefont {H.}~\bibnamefont {{R{\`e}me}}}, \ and\ \bibinfo {author}
  {\bibfnamefont {L.}~\bibnamefont {{Kistler}}},\ }\href@noop {} {\bibfield
  {journal} {\bibinfo  {journal} {Geophys. Res. Lett.},\ }\textbf {\bibinfo
  {volume} {30}},\ \bibinfo {pages} {1579} (\bibinfo {year}
  {2003})}\BibitemShut {NoStop}%
\bibitem [{\citenamefont {Eastwood}\ \emph {et~al.}(2010)\citenamefont
  {Eastwood}, \citenamefont {Phan}, \citenamefont {{\O}ieroset},\ and\
  \citenamefont {Shay}}]{eastwood:2010}%
  \BibitemOpen
  \bibfield  {author} {\bibinfo {author} {\bibfnamefont {J.~P.}\ \bibnamefont
  {Eastwood}}, \bibinfo {author} {\bibfnamefont {T.~D.}\ \bibnamefont {Phan}},
  \bibinfo {author} {\bibfnamefont {M.}~\bibnamefont {{\O}ieroset}}, \ and\
  \bibinfo {author} {\bibfnamefont {M.}~\bibnamefont {Shay}},\ }\href@noop {}
  {\bibfield  {journal} {\bibinfo  {journal} {J. Geophys. Res.},\ }\textbf
  {\bibinfo {volume} {115}},\ \bibinfo {eid} {A08215} (\bibinfo {year}
  {2010})}\BibitemShut {NoStop}%
\bibitem [{\citenamefont {{Birn}}\ \emph {et~al.}(1996)\citenamefont {{Birn}},
  \citenamefont {{Hesse}},\ and\ \citenamefont {{Schindler}}}]{birn:1996}%
  \BibitemOpen
  \bibfield  {author} {\bibinfo {author} {\bibfnamefont {J.}~\bibnamefont
  {{Birn}}}, \bibinfo {author} {\bibfnamefont {M.}~\bibnamefont {{Hesse}}}, \
  and\ \bibinfo {author} {\bibfnamefont {K.}~\bibnamefont {{Schindler}}},\
  }\href@noop {} {\bibfield  {journal} {\bibinfo  {journal} {J. Geophys.
  Res.},\ }\textbf {\bibinfo {volume} {101}},\ \bibinfo {pages} {12939}
  (\bibinfo {year} {1996})}\BibitemShut {NoStop}%
\bibitem [{\citenamefont {{Ohtani}}\ and\ \citenamefont
  {{Raeder}}(2004)}]{ohtani:2004}%
  \BibitemOpen
  \bibfield  {author} {\bibinfo {author} {\bibfnamefont {S.}~\bibnamefont
  {{Ohtani}}}\ and\ \bibinfo {author} {\bibfnamefont {J.}~\bibnamefont
  {{Raeder}}},\ }\href@noop {} {\bibfield  {journal} {\bibinfo  {journal} {J.
  Geophys. Res.},\ }\textbf {\bibinfo {volume} {109}},\ \bibinfo {pages} {1207}
  (\bibinfo {year} {2004})}\BibitemShut {NoStop}%
\bibitem [{\citenamefont {{Kuznetsova}}\ \emph {et~al.}(2007)\citenamefont
  {{Kuznetsova}}, \citenamefont {{Hesse}}, \citenamefont {{Rast{\"a}tter}},
  \citenamefont {{Taktakishvili}}, \citenamefont {{Toth}}, \citenamefont {{De
  Zeeuw}}, \citenamefont {{Ridley}},\ and\ \citenamefont
  {{Gombosi}}}]{kuznetsova:2007}%
  \BibitemOpen
  \bibfield  {author} {\bibinfo {author} {\bibfnamefont {M.~M.}\ \bibnamefont
  {{Kuznetsova}}}, \bibinfo {author} {\bibfnamefont {M.}~\bibnamefont
  {{Hesse}}}, \bibinfo {author} {\bibfnamefont {L.}~\bibnamefont
  {{Rast{\"a}tter}}}, \bibinfo {author} {\bibfnamefont {A.}~\bibnamefont
  {{Taktakishvili}}}, \bibinfo {author} {\bibfnamefont {G.}~\bibnamefont
  {{Toth}}}, \bibinfo {author} {\bibfnamefont {D.~L.}\ \bibnamefont {{De
  Zeeuw}}}, \bibinfo {author} {\bibfnamefont {A.}~\bibnamefont {{Ridley}}}, \
  and\ \bibinfo {author} {\bibfnamefont {T.~I.}\ \bibnamefont {{Gombosi}}},\
  }\href@noop {} {\bibfield  {journal} {\bibinfo  {journal} {J. Geophys.
  Res.},\ }\textbf {\bibinfo {volume} {112}},\ \bibinfo {eid} {10210} (\bibinfo
  {year} {2007})}\BibitemShut {NoStop}%
\bibitem [{\citenamefont {{Zhu}}\ \emph {et~al.}(2009)\citenamefont {{Zhu}},
  \citenamefont {{Raeder}}, \citenamefont {{Germaschewski}},\ and\
  \citenamefont {{Hegna}}}]{zhu:2009}%
  \BibitemOpen
  \bibfield  {author} {\bibinfo {author} {\bibfnamefont {P.}~\bibnamefont
  {{Zhu}}}, \bibinfo {author} {\bibfnamefont {J.}~\bibnamefont {{Raeder}}},
  \bibinfo {author} {\bibfnamefont {K.}~\bibnamefont {{Germaschewski}}}, \ and\
  \bibinfo {author} {\bibfnamefont {C.~C.}\ \bibnamefont {{Hegna}}},\
  }\href@noop {} {\bibfield  {journal} {\bibinfo  {journal} {Ann. Geophys.},\
  }\textbf {\bibinfo {volume} {27}},\ \bibinfo {pages} {1129} (\bibinfo {year}
  {2009})}\BibitemShut {NoStop}%
\bibitem [{\citenamefont {{Owen}}\ and\ \citenamefont
  {{Cowley}}(1987){\natexlab{a}}}]{owen:1987:451}%
  \BibitemOpen
  \bibfield  {author} {\bibinfo {author} {\bibfnamefont {C.~J.}\ \bibnamefont
  {{Owen}}}\ and\ \bibinfo {author} {\bibfnamefont {S.~W.~H.}\ \bibnamefont
  {{Cowley}}},\ }\href@noop {} {\bibfield  {journal} {\bibinfo  {journal}
  {Planet. Space Sci.},\ }\textbf {\bibinfo {volume} {35}},\ \bibinfo {pages}
  {451} (\bibinfo {year} {1987}{\natexlab{a}})}\BibitemShut {NoStop}%
\bibitem [{\citenamefont {{Owen}}\ and\ \citenamefont
  {{Cowley}}(1987){\natexlab{b}}}]{owen:1987:467}%
  \BibitemOpen
  \bibfield  {author} {\bibinfo {author} {\bibfnamefont {C.~J.}\ \bibnamefont
  {{Owen}}}\ and\ \bibinfo {author} {\bibfnamefont {S.~W.~H.}\ \bibnamefont
  {{Cowley}}},\ }\href@noop {} {\bibfield  {journal} {\bibinfo  {journal}
  {Planet. Space Sci.},\ }\textbf {\bibinfo {volume} {35}},\ \bibinfo {pages}
  {467} (\bibinfo {year} {1987}{\natexlab{b}})}\BibitemShut {NoStop}%
\bibitem [{\citenamefont {{Kiehas}}\ \emph {et~al.}(2007)\citenamefont
  {{Kiehas}}, \citenamefont {{Semenov}}, \citenamefont {{Kubyshkin}},
  \citenamefont {{Tolstykh}}, \citenamefont {{Penz}},\ and\ \citenamefont
  {{Biernat}}}]{kiehas:2007}%
  \BibitemOpen
  \bibfield  {author} {\bibinfo {author} {\bibfnamefont {S.~A.}\ \bibnamefont
  {{Kiehas}}}, \bibinfo {author} {\bibfnamefont {V.~S.}\ \bibnamefont
  {{Semenov}}}, \bibinfo {author} {\bibfnamefont {I.~V.}\ \bibnamefont
  {{Kubyshkin}}}, \bibinfo {author} {\bibfnamefont {Y.~V.}\ \bibnamefont
  {{Tolstykh}}}, \bibinfo {author} {\bibfnamefont {T.}~\bibnamefont {{Penz}}},
  \ and\ \bibinfo {author} {\bibfnamefont {H.~K.}\ \bibnamefont {{Biernat}}},\
  }\href@noop {} {\bibfield  {journal} {\bibinfo  {journal} {Ann. Geophys.},\
  }\textbf {\bibinfo {volume} {25}},\ \bibinfo {pages} {293} (\bibinfo {year}
  {2007})}\BibitemShut {NoStop}%
\bibitem [{\citenamefont {Kiehas}\ \emph {et~al.}(2009)\citenamefont {Kiehas},
  \citenamefont {Semenov}, \citenamefont {Kubyshkina}, \citenamefont
  {Angelopoulos}, \citenamefont {Nakamura}, \citenamefont {Keika},
  \citenamefont {Ivanova}, \citenamefont {Biernat}, \citenamefont {Baumjohann},
  \citenamefont {Mende}, \citenamefont {Magnes}, \citenamefont {Auster},
  \citenamefont {Fornacon}, \citenamefont {Larson}, \citenamefont {Carlson},
  \citenamefont {Bonnell},\ and\ \citenamefont {McFadden}}]{kiehas:2009}%
  \BibitemOpen
  \bibfield  {author} {\bibinfo {author} {\bibfnamefont {S.}~\bibnamefont
  {Kiehas}}, \bibinfo {author} {\bibfnamefont {V.~S.}\ \bibnamefont {Semenov}},
  \bibinfo {author} {\bibfnamefont {M.}~\bibnamefont {Kubyshkina}}, \bibinfo
  {author} {\bibfnamefont {V.}~\bibnamefont {Angelopoulos}}, \bibinfo {author}
  {\bibfnamefont {R.}~\bibnamefont {Nakamura}}, \bibinfo {author}
  {\bibfnamefont {K.}~\bibnamefont {Keika}}, \bibinfo {author} {\bibfnamefont
  {I.}~\bibnamefont {Ivanova}}, \bibinfo {author} {\bibfnamefont {H.~K.}\
  \bibnamefont {Biernat}}, \bibinfo {author} {\bibfnamefont {W.}~\bibnamefont
  {Baumjohann}}, \bibinfo {author} {\bibfnamefont {S.~B.}\ \bibnamefont
  {Mende}}, \bibinfo {author} {\bibfnamefont {W.}~\bibnamefont {Magnes}},
  \bibinfo {author} {\bibfnamefont {U.}~\bibnamefont {Auster}}, \bibinfo
  {author} {\bibfnamefont {K.-H.}\ \bibnamefont {Fornacon}}, \bibinfo {author}
  {\bibfnamefont {D.}~\bibnamefont {Larson}}, \bibinfo {author} {\bibfnamefont
  {C.~W.}\ \bibnamefont {Carlson}}, \bibinfo {author} {\bibfnamefont
  {J.}~\bibnamefont {Bonnell}}, \ and\ \bibinfo {author} {\bibfnamefont
  {J.}~\bibnamefont {McFadden}},\ }\href@noop {} {\bibfield  {journal}
  {\bibinfo  {journal} {J. Geophys. Res.},\ }\textbf {\bibinfo {volume}
  {114}},\ \bibinfo {eid} {A00C20} (\bibinfo {year} {2009})}\BibitemShut
  {NoStop}%
\bibitem [{\citenamefont {{Kopp}}\ and\ \citenamefont
  {{Pneuman}}(1976)}]{kopp:pneuman:1976}%
  \BibitemOpen
  \bibfield  {author} {\bibinfo {author} {\bibfnamefont {R.~A.}\ \bibnamefont
  {{Kopp}}}\ and\ \bibinfo {author} {\bibfnamefont {G.~W.}\ \bibnamefont
  {{Pneuman}}},\ }\href@noop {} {\bibfield  {journal} {\bibinfo  {journal}
  {Sol. Phys.},\ }\textbf {\bibinfo {volume} {50}},\ \bibinfo {pages} {85}
  (\bibinfo {year} {1976})}\BibitemShut {NoStop}%
\bibitem [{\citenamefont {{Forbes}}\ and\ \citenamefont
  {{Acton}}(1996)}]{forbes:acton:1996}%
  \BibitemOpen
  \bibfield  {author} {\bibinfo {author} {\bibfnamefont {T.~G.}\ \bibnamefont
  {{Forbes}}}\ and\ \bibinfo {author} {\bibfnamefont {L.~W.}\ \bibnamefont
  {{Acton}}},\ }\href@noop {} {\bibfield  {journal} {\bibinfo  {journal}
  {Astrophys. J.},\ }\textbf {\bibinfo {volume} {459}},\ \bibinfo {pages} {330}
  (\bibinfo {year} {1996})}\BibitemShut {NoStop}%
\bibitem [{\citenamefont {{Lin}}\ and\ \citenamefont
  {{Forbes}}(2000)}]{linforbes:2000}%
  \BibitemOpen
  \bibfield  {author} {\bibinfo {author} {\bibfnamefont {J.}~\bibnamefont
  {{Lin}}}\ and\ \bibinfo {author} {\bibfnamefont {T.~G.}\ \bibnamefont
  {{Forbes}}},\ }\href@noop {} {\bibfield  {journal} {\bibinfo  {journal} {J.
  Geophys. Res.},\ }\textbf {\bibinfo {volume} {105}},\ \bibinfo {pages} {2375}
  (\bibinfo {year} {2000})}\BibitemShut {NoStop}%
\bibitem [{\citenamefont {{Ciaravella}}\ \emph {et~al.}(2002)\citenamefont
  {{Ciaravella}}, \citenamefont {{Raymond}}, \citenamefont {{Li}},
  \citenamefont {{Reiser}}, \citenamefont {{Gardner}}, \citenamefont {{Ko}},\
  and\ \citenamefont {{Fineschi}}}]{2002ApJ...575.1116C}%
  \BibitemOpen
  \bibfield  {author} {\bibinfo {author} {\bibfnamefont {A.}~\bibnamefont
  {{Ciaravella}}}, \bibinfo {author} {\bibfnamefont {J.~C.}\ \bibnamefont
  {{Raymond}}}, \bibinfo {author} {\bibfnamefont {J.}~\bibnamefont {{Li}}},
  \bibinfo {author} {\bibfnamefont {P.}~\bibnamefont {{Reiser}}}, \bibinfo
  {author} {\bibfnamefont {L.~D.}\ \bibnamefont {{Gardner}}}, \bibinfo {author}
  {\bibfnamefont {Y.}~\bibnamefont {{Ko}}}, \ and\ \bibinfo {author}
  {\bibfnamefont {S.}~\bibnamefont {{Fineschi}}},\ }\href@noop {} {\bibfield
  {journal} {\bibinfo  {journal} {Astrophys. J.},\ }\textbf {\bibinfo {volume}
  {575}},\ \bibinfo {pages} {1116} (\bibinfo {year} {2002})}\BibitemShut
  {NoStop}%
\bibitem [{\citenamefont {{Ciaravella}}\ and\ \citenamefont
  {{Raymond}}(2008)}]{2008ApJ...686.1372C}%
  \BibitemOpen
  \bibfield  {author} {\bibinfo {author} {\bibfnamefont {A.}~\bibnamefont
  {{Ciaravella}}}\ and\ \bibinfo {author} {\bibfnamefont {J.~C.}\ \bibnamefont
  {{Raymond}}},\ }\href@noop {} {\bibfield  {journal} {\bibinfo  {journal}
  {Astrophys. J.},\ }\textbf {\bibinfo {volume} {686}},\ \bibinfo {pages}
  {1372} (\bibinfo {year} {2008})}\BibitemShut {NoStop}%
\bibitem [{\citenamefont {{Schettino}}\ \emph {et~al.}(2010)\citenamefont
  {{Schettino}}, \citenamefont {{Poletto}},\ and\ \citenamefont
  {{Romoli}}}]{2010ApJ...708.1135S}%
  \BibitemOpen
  \bibfield  {author} {\bibinfo {author} {\bibfnamefont {G.}~\bibnamefont
  {{Schettino}}}, \bibinfo {author} {\bibfnamefont {G.}~\bibnamefont
  {{Poletto}}}, \ and\ \bibinfo {author} {\bibfnamefont {M.}~\bibnamefont
  {{Romoli}}},\ }\href@noop {} {\bibfield  {journal} {\bibinfo  {journal}
  {Astrophys. J.},\ }\textbf {\bibinfo {volume} {708}},\ \bibinfo {pages}
  {1135} (\bibinfo {year} {2010})}\BibitemShut {NoStop}%
\bibitem [{\citenamefont {{Savage}}\ \emph {et~al.}(2010)\citenamefont
  {{Savage}}, \citenamefont {{McKenzie}}, \citenamefont {{Reeves}},
  \citenamefont {{Forbes}},\ and\ \citenamefont {{Longcope}}}]{savage:2010}%
  \BibitemOpen
  \bibfield  {author} {\bibinfo {author} {\bibfnamefont {S.~L.}\ \bibnamefont
  {{Savage}}}, \bibinfo {author} {\bibfnamefont {D.~E.}\ \bibnamefont
  {{McKenzie}}}, \bibinfo {author} {\bibfnamefont {K.~K.}\ \bibnamefont
  {{Reeves}}}, \bibinfo {author} {\bibfnamefont {T.~G.}\ \bibnamefont
  {{Forbes}}}, \ and\ \bibinfo {author} {\bibfnamefont {D.~W.}\ \bibnamefont
  {{Longcope}}},\ }\bibfield  {title} {\enquote {\bibinfo {title}
  {{Reconnection Outflows and Current Sheet Observed with Hinode/XRT in the
  2008 April 9 ``Cartwheel CME'' Flare}},}\ }\href@noop {} {\bibfield
  {journal} {\bibinfo  {journal} {Astrophys.\ J., in press},\ }\textbf
  {\bibinfo {volume} {721}} (\bibinfo {year} {2010})}\BibitemShut {NoStop}%
\bibitem [{\citenamefont {{Linker}}\ \emph {et~al.}(2003)\citenamefont
  {{Linker}}, \citenamefont {{Miki{\'c}}}, \citenamefont {{Lionello}},
  \citenamefont {{Riley}}, \citenamefont {{Amari}},\ and\ \citenamefont
  {{Odstrcil}}}]{linker:2003}%
  \BibitemOpen
  \bibfield  {author} {\bibinfo {author} {\bibfnamefont {J.~A.}\ \bibnamefont
  {{Linker}}}, \bibinfo {author} {\bibfnamefont {Z.}~\bibnamefont
  {{Miki{\'c}}}}, \bibinfo {author} {\bibfnamefont {R.}~\bibnamefont
  {{Lionello}}}, \bibinfo {author} {\bibfnamefont {P.}~\bibnamefont {{Riley}}},
  \bibinfo {author} {\bibfnamefont {T.}~\bibnamefont {{Amari}}}, \ and\
  \bibinfo {author} {\bibfnamefont {D.}~\bibnamefont {{Odstrcil}}},\
  }\href@noop {} {\bibfield  {journal} {\bibinfo  {journal} {Phys. Plasmas},\
  }\textbf {\bibinfo {volume} {10}},\ \bibinfo {pages} {1971} (\bibinfo {year}
  {2003})}\BibitemShut {NoStop}%
\bibitem [{\citenamefont {Reeves}\ \emph {et~al.}(2010)\citenamefont {Reeves},
  \citenamefont {Linker}, \citenamefont {Miki{\'c}},\ and\ \citenamefont
  {Forbes}}]{reeves:2010}%
  \BibitemOpen
  \bibfield  {author} {\bibinfo {author} {\bibfnamefont {K.~K.}\ \bibnamefont
  {Reeves}}, \bibinfo {author} {\bibfnamefont {J.~A.}\ \bibnamefont {Linker}},
  \bibinfo {author} {\bibfnamefont {Z.}~\bibnamefont {Miki{\'c}}}, \ and\
  \bibinfo {author} {\bibfnamefont {T.~G.}\ \bibnamefont {Forbes}},\
  }\href@noop {} {\bibfield  {journal} {\bibinfo  {journal} {Astrophys. J.},\
  }\textbf {\bibinfo {volume} {721}},\ \bibinfo {pages} {1547} (\bibinfo {year}
  {2010})}\BibitemShut {NoStop}%
\bibitem [{\citenamefont {{Yamada}}\ \emph {et~al.}(1997)\citenamefont
  {{Yamada}}, \citenamefont {{Ji}}, \citenamefont {{Hsu}}, \citenamefont
  {{Carter}}, \citenamefont {{Kulsrud}}, \citenamefont {{Bretz}}, \citenamefont
  {{Jobes}}, \citenamefont {{Ono}},\ and\ \citenamefont
  {{Perkins}}}]{yamada:mrx}%
  \BibitemOpen
  \bibfield  {author} {\bibinfo {author} {\bibfnamefont {M.}~\bibnamefont
  {{Yamada}}}, \bibinfo {author} {\bibfnamefont {H.}~\bibnamefont {{Ji}}},
  \bibinfo {author} {\bibfnamefont {S.}~\bibnamefont {{Hsu}}}, \bibinfo
  {author} {\bibfnamefont {T.}~\bibnamefont {{Carter}}}, \bibinfo {author}
  {\bibfnamefont {R.}~\bibnamefont {{Kulsrud}}}, \bibinfo {author}
  {\bibfnamefont {N.}~\bibnamefont {{Bretz}}}, \bibinfo {author} {\bibfnamefont
  {F.}~\bibnamefont {{Jobes}}}, \bibinfo {author} {\bibfnamefont
  {Y.}~\bibnamefont {{Ono}}}, \ and\ \bibinfo {author} {\bibfnamefont
  {F.}~\bibnamefont {{Perkins}}},\ }\href@noop {} {\bibfield  {journal}
  {\bibinfo  {journal} {Phys. Plasmas},\ }\textbf {\bibinfo {volume} {4}},\
  \bibinfo {pages} {1936} (\bibinfo {year} {1997})}\BibitemShut {NoStop}%
\bibitem [{\citenamefont {{Inomoto}}\ \emph {et~al.}(2006)\citenamefont
  {{Inomoto}}, \citenamefont {{Gerhardt}}, \citenamefont {{Yamada}},
  \citenamefont {{Ji}}, \citenamefont {{Belova}}, \citenamefont {{Kuritsyn}},\
  and\ \citenamefont {{Ren}}}]{inomoto:counter}%
  \BibitemOpen
  \bibfield  {author} {\bibinfo {author} {\bibfnamefont {M.}~\bibnamefont
  {{Inomoto}}}, \bibinfo {author} {\bibfnamefont {S.~P.}\ \bibnamefont
  {{Gerhardt}}}, \bibinfo {author} {\bibfnamefont {M.}~\bibnamefont
  {{Yamada}}}, \bibinfo {author} {\bibfnamefont {H.}~\bibnamefont {{Ji}}},
  \bibinfo {author} {\bibfnamefont {E.}~\bibnamefont {{Belova}}}, \bibinfo
  {author} {\bibfnamefont {A.}~\bibnamefont {{Kuritsyn}}}, \ and\ \bibinfo
  {author} {\bibfnamefont {Y.}~\bibnamefont {{Ren}}},\ }\href@noop {}
  {\bibfield  {journal} {\bibinfo  {journal} {Phys. Rev. Lett.},\ }\textbf
  {\bibinfo {volume} {97}},\ \bibinfo {eid} {135002} (\bibinfo {year}
  {2006})}\BibitemShut {NoStop}%
\bibitem [{\citenamefont {Murphy}\ and\ \citenamefont
  {Sovinec}(2008)}]{murphy:mrx}%
  \BibitemOpen
  \bibfield  {author} {\bibinfo {author} {\bibfnamefont {N.~A.}\ \bibnamefont
  {Murphy}}\ and\ \bibinfo {author} {\bibfnamefont {C.~R.}\ \bibnamefont
  {Sovinec}},\ }\href@noop {} {\bibfield  {journal} {\bibinfo  {journal} {Phys.
  Plasmas},\ }\textbf {\bibinfo {volume} {15}},\ \bibinfo {eid} {042313}
  (\bibinfo {year} {2008})}\BibitemShut {NoStop}%
\bibitem [{\citenamefont {{Ono}}\ \emph
  {et~al.}(1993){\natexlab{a}}\citenamefont {{Ono}}, \citenamefont {{Morita}},\
  and\ \citenamefont {Katsurai}}]{ono:1993}%
  \BibitemOpen
  \bibfield  {author} {\bibinfo {author} {\bibfnamefont {Y.}~\bibnamefont
  {{Ono}}}, \bibinfo {author} {\bibfnamefont {A.}~\bibnamefont {{Morita}}}, \
  and\ \bibinfo {author} {\bibfnamefont {M.}~\bibnamefont {Katsurai}},\
  }\href@noop {} {\bibfield  {journal} {\bibinfo  {journal} {Phys. Fluids B},\
  }\textbf {\bibinfo {volume} {5}},\ \bibinfo {pages} {3691} (\bibinfo {year}
  {1993}{\natexlab{a}})}\BibitemShut {NoStop}%
\bibitem [{\citenamefont {{Ono}}\ \emph
  {et~al.}(1993){\natexlab{b}}\citenamefont {{Ono}}, \citenamefont {Inomoto},
  \citenamefont {Okazaki},\ and\ \citenamefont {Ueda}}]{ono:1997}%
  \BibitemOpen
  \bibfield  {author} {\bibinfo {author} {\bibfnamefont {Y.}~\bibnamefont
  {{Ono}}}, \bibinfo {author} {\bibfnamefont {M.}~\bibnamefont {Inomoto}},
  \bibinfo {author} {\bibfnamefont {T.}~\bibnamefont {Okazaki}}, \ and\
  \bibinfo {author} {\bibfnamefont {Y.}~\bibnamefont {Ueda}},\ }\href@noop {}
  {\bibfield  {journal} {\bibinfo  {journal} {Phys. Plasmas},\ }\textbf
  {\bibinfo {volume} {4}},\ \bibinfo {pages} {1953} (\bibinfo {year}
  {1993}{\natexlab{b}})}\BibitemShut {NoStop}%
\bibitem [{\citenamefont {{Rogers}}\ and\ \citenamefont
  {{Zakharov}}(1995)}]{rogers:1995}%
  \BibitemOpen
  \bibfield  {author} {\bibinfo {author} {\bibfnamefont {B.}~\bibnamefont
  {{Rogers}}}\ and\ \bibinfo {author} {\bibfnamefont {L.}~\bibnamefont
  {{Zakharov}}},\ }\href@noop {} {\bibfield  {journal} {\bibinfo  {journal}
  {Phys. Plasmas},\ }\textbf {\bibinfo {volume} {2}},\ \bibinfo {pages} {3420}
  (\bibinfo {year} {1995})}\BibitemShut {NoStop}%
\bibitem [{\citenamefont {{Swisdak}}\ \emph {et~al.}(2003)\citenamefont
  {{Swisdak}}, \citenamefont {{Rogers}}, \citenamefont {{Drake}},\ and\
  \citenamefont {{Shay}}}]{swisdak:diamagnetic}%
  \BibitemOpen
  \bibfield  {author} {\bibinfo {author} {\bibfnamefont {M.}~\bibnamefont
  {{Swisdak}}}, \bibinfo {author} {\bibfnamefont {B.~N.}\ \bibnamefont
  {{Rogers}}}, \bibinfo {author} {\bibfnamefont {J.~F.}\ \bibnamefont
  {{Drake}}}, \ and\ \bibinfo {author} {\bibfnamefont {M.~A.}\ \bibnamefont
  {{Shay}}},\ }\href@noop {} {\bibfield  {journal} {\bibinfo  {journal} {J.
  Geophys. Res.},\ }\textbf {\bibinfo {volume} {108}},\ \bibinfo {pages} {23}
  (\bibinfo {year} {2003})}\BibitemShut {NoStop}%
\bibitem [{\citenamefont {{Phan}}\ \emph {et~al.}(2010)\citenamefont {{Phan}},
  \citenamefont {{Gosling}}, \citenamefont {{Paschmann}}, \citenamefont
  {{Pasma}}, \citenamefont {{Drake}}, \citenamefont {{{\O}ieroset}},
  \citenamefont {{Larson}}, \citenamefont {{Lin}},\ and\ \citenamefont
  {{Davis}}}]{phan:2010}%
  \BibitemOpen
  \bibfield  {author} {\bibinfo {author} {\bibfnamefont {T.~D.}\ \bibnamefont
  {{Phan}}}, \bibinfo {author} {\bibfnamefont {J.~T.}\ \bibnamefont
  {{Gosling}}}, \bibinfo {author} {\bibfnamefont {G.}~\bibnamefont
  {{Paschmann}}}, \bibinfo {author} {\bibfnamefont {C.}~\bibnamefont
  {{Pasma}}}, \bibinfo {author} {\bibfnamefont {J.~F.}\ \bibnamefont
  {{Drake}}}, \bibinfo {author} {\bibfnamefont {M.}~\bibnamefont
  {{{\O}ieroset}}}, \bibinfo {author} {\bibfnamefont {D.}~\bibnamefont
  {{Larson}}}, \bibinfo {author} {\bibfnamefont {R.~P.}\ \bibnamefont {{Lin}}},
  \ and\ \bibinfo {author} {\bibfnamefont {M.~S.}\ \bibnamefont {{Davis}}},\
  }\href@noop {} {\bibfield  {journal} {\bibinfo  {journal} {Astrophys. J.
  Lett.},\ }\textbf {\bibinfo {volume} {719}},\ \bibinfo {pages} {L199}
  (\bibinfo {year} {2010})}\BibitemShut {NoStop}%
\bibitem [{\citenamefont {Oka}\ \emph {et~al.}(2008)\citenamefont {Oka},
  \citenamefont {Fujimoto}, \citenamefont {Nakamura}, \citenamefont
  {Shinohara},\ and\ \citenamefont {Nishikawa}}]{oka:asym}%
  \BibitemOpen
  \bibfield  {author} {\bibinfo {author} {\bibfnamefont {M.}~\bibnamefont
  {Oka}}, \bibinfo {author} {\bibfnamefont {M.}~\bibnamefont {Fujimoto}},
  \bibinfo {author} {\bibfnamefont {T.~K.~M.}\ \bibnamefont {Nakamura}},
  \bibinfo {author} {\bibfnamefont {I.}~\bibnamefont {Shinohara}}, \ and\
  \bibinfo {author} {\bibfnamefont {K.}~\bibnamefont {Nishikawa}},\ }\href@noop
  {} {\bibfield  {journal} {\bibinfo  {journal} {Phys. Rev. Lett.},\ }\textbf
  {\bibinfo {volume} {101}},\ \bibinfo {eid} {205004} (\bibinfo {year}
  {2008})}\BibitemShut {NoStop}%
\bibitem [{\citenamefont {{Sovinec}}\ \emph {et~al.}(2004)\citenamefont
  {{Sovinec}}, \citenamefont {{Glasser}}, \citenamefont {{Gianakon}},
  \citenamefont {{Barnes}}, \citenamefont {{Nebel}}, \citenamefont {{Kruger}},
  \citenamefont {{Schnack}}, \citenamefont {{Plimpton}}, \citenamefont
  {{Tarditi}},\ and\ \citenamefont {{Chu}}}]{sovinec:jcp}%
  \BibitemOpen
  \bibfield  {author} {\bibinfo {author} {\bibfnamefont {C.~R.}\ \bibnamefont
  {{Sovinec}}}, \bibinfo {author} {\bibfnamefont {A.~H.}\ \bibnamefont
  {{Glasser}}}, \bibinfo {author} {\bibfnamefont {T.~A.}\ \bibnamefont
  {{Gianakon}}}, \bibinfo {author} {\bibfnamefont {D.~C.}\ \bibnamefont
  {{Barnes}}}, \bibinfo {author} {\bibfnamefont {R.~A.}\ \bibnamefont
  {{Nebel}}}, \bibinfo {author} {\bibfnamefont {S.~E.}\ \bibnamefont
  {{Kruger}}}, \bibinfo {author} {\bibfnamefont {D.~D.}\ \bibnamefont
  {{Schnack}}}, \bibinfo {author} {\bibfnamefont {S.~J.}\ \bibnamefont
  {{Plimpton}}}, \bibinfo {author} {\bibfnamefont {A.}~\bibnamefont
  {{Tarditi}}}, \ and\ \bibinfo {author} {\bibfnamefont {M.~S.}\ \bibnamefont
  {{Chu}}},\ }\href@noop {} {\bibfield  {journal} {\bibinfo  {journal} {J.
  Comput. Phys.},\ }\textbf {\bibinfo {volume} {195}},\ \bibinfo {pages} {355}
  (\bibinfo {year} {2004})}\BibitemShut {NoStop}%
\bibitem [{\citenamefont {{Sovinec}}\ \emph {et~al.}(2005)\citenamefont
  {{Sovinec}}, \citenamefont {{Schnack}}, \citenamefont {{Pankin}},
  \citenamefont {{Brennan}}, \citenamefont {{Tian}}, \citenamefont {{Barnes}},
  \citenamefont {{Kruger}}, \citenamefont {{Held}}, \citenamefont {{Kim}},
  \citenamefont {{Li}}, \citenamefont {{Kaushik}}, \citenamefont {{Jardin}},\
  and\ \citenamefont {{the NIMROD Team}}}]{sovinec:jop}%
  \BibitemOpen
  \bibfield  {author} {\bibinfo {author} {\bibfnamefont {C.~R.}\ \bibnamefont
  {{Sovinec}}}, \bibinfo {author} {\bibfnamefont {D.~D.}\ \bibnamefont
  {{Schnack}}}, \bibinfo {author} {\bibfnamefont {A.~Y.}\ \bibnamefont
  {{Pankin}}}, \bibinfo {author} {\bibfnamefont {D.~P.}\ \bibnamefont
  {{Brennan}}}, \bibinfo {author} {\bibfnamefont {H.}~\bibnamefont {{Tian}}},
  \bibinfo {author} {\bibfnamefont {D.~C.}\ \bibnamefont {{Barnes}}}, \bibinfo
  {author} {\bibfnamefont {S.~E.}\ \bibnamefont {{Kruger}}}, \bibinfo {author}
  {\bibfnamefont {E.~D.}\ \bibnamefont {{Held}}}, \bibinfo {author}
  {\bibfnamefont {C.~C.}\ \bibnamefont {{Kim}}}, \bibinfo {author}
  {\bibfnamefont {X.~S.}\ \bibnamefont {{Li}}}, \bibinfo {author}
  {\bibfnamefont {D.~K.}\ \bibnamefont {{Kaushik}}}, \bibinfo {author}
  {\bibfnamefont {S.~C.}\ \bibnamefont {{Jardin}}}, \ and\ \bibinfo {author}
  {\bibnamefont {{the NIMROD Team}}},\ }\href@noop {} {\bibfield  {journal}
  {\bibinfo  {journal} {J. Phys.: Conf. Ser.},\ }\textbf {\bibinfo {volume}
  {16}},\ \bibinfo {pages} {25} (\bibinfo {year} {2005})}\BibitemShut {NoStop}%
\bibitem [{\citenamefont {Sovinec}\ and\ \citenamefont
  {King}(2010)}]{sovinec:2009}%
  \BibitemOpen
  \bibfield  {author} {\bibinfo {author} {\bibfnamefont {C.~R.}\ \bibnamefont
  {Sovinec}}\ and\ \bibinfo {author} {\bibfnamefont {J.~R.}\ \bibnamefont
  {King}},\ }\href@noop {} {\bibfield  {journal} {\bibinfo  {journal} {J.
  Comput. Phys.},\ }\textbf {\bibinfo {volume} {229}},\ \bibinfo {pages} {5803
  } (\bibinfo {year} {2010})}\BibitemShut {NoStop}%
\bibitem [{\citenamefont {{Hooper}}\ \emph {et~al.}(2005)\citenamefont
  {{Hooper}}, \citenamefont {{Kopriva}}, \citenamefont {{Cohen}}, \citenamefont
  {{Hill}}, \citenamefont {{McLean}}, \citenamefont {{Wood}}, \citenamefont
  {{Woodruff}},\ and\ \citenamefont {{Sovinec}}}]{hooper:recon}%
  \BibitemOpen
  \bibfield  {author} {\bibinfo {author} {\bibfnamefont {E.~B.}\ \bibnamefont
  {{Hooper}}}, \bibinfo {author} {\bibfnamefont {T.~A.}\ \bibnamefont
  {{Kopriva}}}, \bibinfo {author} {\bibfnamefont {B.~I.}\ \bibnamefont
  {{Cohen}}}, \bibinfo {author} {\bibfnamefont {D.~N.}\ \bibnamefont {{Hill}}},
  \bibinfo {author} {\bibfnamefont {H.~S.}\ \bibnamefont {{McLean}}}, \bibinfo
  {author} {\bibfnamefont {R.~D.}\ \bibnamefont {{Wood}}}, \bibinfo {author}
  {\bibfnamefont {S.}~\bibnamefont {{Woodruff}}}, \ and\ \bibinfo {author}
  {\bibfnamefont {C.~R.}\ \bibnamefont {{Sovinec}}},\ }\href@noop {} {\bibfield
   {journal} {\bibinfo  {journal} {Phys. Plasmas},\ }\textbf {\bibinfo {volume}
  {12}},\ \bibinfo {eid} {092503} (\bibinfo {year} {2005})}\BibitemShut
  {NoStop}%
\bibitem [{\citenamefont {{Nakamura}}\ \emph {et~al.}(2010)\citenamefont
  {{Nakamura}}, \citenamefont {{Fujimoto}},\ and\ \citenamefont
  {{Sekiya}}}]{nakamura:2010}%
  \BibitemOpen
  \bibfield  {author} {\bibinfo {author} {\bibfnamefont {T.~K.~M.}\
  \bibnamefont {{Nakamura}}}, \bibinfo {author} {\bibfnamefont
  {M.}~\bibnamefont {{Fujimoto}}}, \ and\ \bibinfo {author} {\bibfnamefont
  {H.}~\bibnamefont {{Sekiya}}},\ }\href@noop {} {\bibfield  {journal}
  {\bibinfo  {journal} {Geophys. Res. Lett.},\ }\textbf {\bibinfo {volume}
  {37}},\ \bibinfo {pages} {2103} (\bibinfo {year} {2010})}\BibitemShut
  {NoStop}%
\bibitem [{\citenamefont {{Roussev}}\ \emph {et~al.}(2001)\citenamefont
  {{Roussev}}, \citenamefont {{Galsgaard}}, \citenamefont {{Erd{\'e}lyi}},\
  and\ \citenamefont {{Doyle}}}]{roussev:2001}%
  \BibitemOpen
  \bibfield  {author} {\bibinfo {author} {\bibfnamefont {I.}~\bibnamefont
  {{Roussev}}}, \bibinfo {author} {\bibfnamefont {K.}~\bibnamefont
  {{Galsgaard}}}, \bibinfo {author} {\bibfnamefont {R.}~\bibnamefont
  {{Erd{\'e}lyi}}}, \ and\ \bibinfo {author} {\bibfnamefont {J.~G.}\
  \bibnamefont {{Doyle}}},\ }\href@noop {} {\bibfield  {journal} {\bibinfo
  {journal} {Astron. Astrophys.},\ }\textbf {\bibinfo {volume} {370}},\
  \bibinfo {pages} {298} (\bibinfo {year} {2001})}\BibitemShut {NoStop}%
\bibitem [{\citenamefont {{Galsgaard}}\ and\ \citenamefont
  {{Roussev}}(2002)}]{galsgaard:2002}%
  \BibitemOpen
  \bibfield  {author} {\bibinfo {author} {\bibfnamefont {K.}~\bibnamefont
  {{Galsgaard}}}\ and\ \bibinfo {author} {\bibfnamefont {I.}~\bibnamefont
  {{Roussev}}},\ }\href@noop {} {\bibfield  {journal} {\bibinfo  {journal}
  {Astron. Astrophys.},\ }\textbf {\bibinfo {volume} {383}},\ \bibinfo {pages}
  {685} (\bibinfo {year} {2002})}\BibitemShut {NoStop}%
\bibitem [{\citenamefont {{Seaton}}(2008)}]{seaton:2008}%
  \BibitemOpen
  \bibfield  {author} {\bibinfo {author} {\bibfnamefont {D.~B.}\ \bibnamefont
  {{Seaton}}},\ }\href@noop {} {Ph.D. thesis},\ \bibinfo  {school} {University
  of New Hampshire} (\bibinfo {year} {2008})\BibitemShut {NoStop}%
\bibitem [{\citenamefont {{Loureiro}}\ \emph {et~al.}(2007)\citenamefont
  {{Loureiro}}, \citenamefont {{Schekochihin}},\ and\ \citenamefont
  {{Cowley}}}]{loureiro:2007}%
  \BibitemOpen
  \bibfield  {author} {\bibinfo {author} {\bibfnamefont {N.~F.}\ \bibnamefont
  {{Loureiro}}}, \bibinfo {author} {\bibfnamefont {A.~A.}\ \bibnamefont
  {{Schekochihin}}}, \ and\ \bibinfo {author} {\bibfnamefont {S.~C.}\
  \bibnamefont {{Cowley}}},\ }\href@noop {} {\bibfield  {journal} {\bibinfo
  {journal} {Phys. Plasmas},\ }\textbf {\bibinfo {volume} {14}},\ \bibinfo
  {eid} {100703} (\bibinfo {year} {2007})}\BibitemShut {NoStop}%
\bibitem [{\citenamefont {{Samtaney}}\ \emph {et~al.}(2009)\citenamefont
  {{Samtaney}}, \citenamefont {{Loureiro}}, \citenamefont {{Uzdensky}},
  \citenamefont {{Schekochihin}},\ and\ \citenamefont
  {{Cowley}}}]{samtaney:2009}%
  \BibitemOpen
  \bibfield  {author} {\bibinfo {author} {\bibfnamefont {R.}~\bibnamefont
  {{Samtaney}}}, \bibinfo {author} {\bibfnamefont {N.~F.}\ \bibnamefont
  {{Loureiro}}}, \bibinfo {author} {\bibfnamefont {D.~A.}\ \bibnamefont
  {{Uzdensky}}}, \bibinfo {author} {\bibfnamefont {A.~A.}\ \bibnamefont
  {{Schekochihin}}}, \ and\ \bibinfo {author} {\bibfnamefont {S.~C.}\
  \bibnamefont {{Cowley}}},\ }\href@noop {} {\bibfield  {journal} {\bibinfo
  {journal} {Phys. Rev. Lett.},\ }\textbf {\bibinfo {volume} {103}},\ \bibinfo
  {eid} {105004} (\bibinfo {year} {2009})}\BibitemShut {NoStop}%
\bibitem [{\citenamefont {{Bhattacharjee}}\ \emph {et~al.}(2009)\citenamefont
  {{Bhattacharjee}}, \citenamefont {{Huang}}, \citenamefont {{Yang}},\ and\
  \citenamefont {{Rogers}}}]{bhattacharjee:2009}%
  \BibitemOpen
  \bibfield  {author} {\bibinfo {author} {\bibfnamefont {A.}~\bibnamefont
  {{Bhattacharjee}}}, \bibinfo {author} {\bibfnamefont {Y.}~\bibnamefont
  {{Huang}}}, \bibinfo {author} {\bibfnamefont {H.}~\bibnamefont {{Yang}}}, \
  and\ \bibinfo {author} {\bibfnamefont {B.}~\bibnamefont {{Rogers}}},\
  }\href@noop {} {\bibfield  {journal} {\bibinfo  {journal} {Phys. Plasmas},\
  }\textbf {\bibinfo {volume} {16}},\ \bibinfo {eid} {112102} (\bibinfo {year}
  {2009})}\BibitemShut {NoStop}%
\bibitem [{\citenamefont {Huang}\ and\ \citenamefont
  {Bhattacharjee}(2010)}]{huang:2010}%
  \BibitemOpen
  \bibfield  {author} {\bibinfo {author} {\bibfnamefont {Y.-M.}\ \bibnamefont
  {Huang}}\ and\ \bibinfo {author} {\bibfnamefont {A.}~\bibnamefont
  {Bhattacharjee}},\ }\href@noop {} {\bibfield  {journal} {\bibinfo  {journal}
  {Phys. Plasmas},\ }\textbf {\bibinfo {volume} {17}},\ \bibinfo {eid} {062104}
  (\bibinfo {year} {2010})}\BibitemShut {NoStop}%
\bibitem [{\citenamefont {{Ni}}\ \emph {et~al.}(2010)\citenamefont {{Ni}},
  \citenamefont {{Germaschewski}}, \citenamefont {{Huang}}, \citenamefont
  {{Sullivan}}, \citenamefont {{Yang}},\ and\ \citenamefont
  {{Bhattacharjee}}}]{ni:2010}%
  \BibitemOpen
  \bibfield  {author} {\bibinfo {author} {\bibfnamefont {L.}~\bibnamefont
  {{Ni}}}, \bibinfo {author} {\bibfnamefont {K.}~\bibnamefont
  {{Germaschewski}}}, \bibinfo {author} {\bibfnamefont {Y.}~\bibnamefont
  {{Huang}}}, \bibinfo {author} {\bibfnamefont {B.~P.}\ \bibnamefont
  {{Sullivan}}}, \bibinfo {author} {\bibfnamefont {H.}~\bibnamefont {{Yang}}},
  \ and\ \bibinfo {author} {\bibfnamefont {A.}~\bibnamefont
  {{Bhattacharjee}}},\ }\href@noop {} {\bibfield  {journal} {\bibinfo
  {journal} {Phys. Plasmas},\ }\textbf {\bibinfo {volume} {17}},\ \bibinfo
  {eid} {052109} (\bibinfo {year} {2010})}\BibitemShut {NoStop}%
\bibitem [{\citenamefont {{Shepherd}}\ and\ \citenamefont
  {{Cassak}}(2010)}]{shepherd:2010}%
  \BibitemOpen
  \bibfield  {author} {\bibinfo {author} {\bibfnamefont {L.~S.}\ \bibnamefont
  {{Shepherd}}}\ and\ \bibinfo {author} {\bibfnamefont {P.~A.}\ \bibnamefont
  {{Cassak}}},\ }\href@noop {} {\bibfield  {journal} {\bibinfo  {journal}
  {Phys. Rev. Lett.},\ }\textbf {\bibinfo {volume} {105}},\ \bibinfo {eid}
  {015004} (\bibinfo {year} {2010})}\BibitemShut {NoStop}%
\bibitem [{\citenamefont {Young}(2010)}]{young:pc:2010}%
  \BibitemOpen
  \bibfield  {author} {\bibinfo {author} {\bibfnamefont {A.~K.}\ \bibnamefont
  {Young}},\ }\href@noop {} {} (\bibinfo {year} {2010}),\ \bibinfo {note}
  {private communication}\BibitemShut {NoStop}%
\bibitem [{\citenamefont {Cassak}\ and\ \citenamefont
  {Shay}(2009)}]{cassak:dissipation}%
  \BibitemOpen
  \bibfield  {author} {\bibinfo {author} {\bibfnamefont {P.~A.}\ \bibnamefont
  {Cassak}}\ and\ \bibinfo {author} {\bibfnamefont {M.~A.}\ \bibnamefont
  {Shay}},\ }\href@noop {} {\bibfield  {journal} {\bibinfo  {journal} {Phys.
  Plasmas},\ }\textbf {\bibinfo {volume} {16}},\ \bibinfo {eid} {{055704}}
  (\bibinfo {year} {2009})}\BibitemShut {NoStop}%
\bibitem [{\citenamefont {{Shibata}}\ and\ \citenamefont
  {{Tanuma}}(2001)}]{shibata:2001}%
  \BibitemOpen
  \bibfield  {author} {\bibinfo {author} {\bibfnamefont {K.}~\bibnamefont
  {{Shibata}}}\ and\ \bibinfo {author} {\bibfnamefont {S.}~\bibnamefont
  {{Tanuma}}},\ }\href@noop {} {\bibfield  {journal} {\bibinfo  {journal}
  {Earth Planets Space},\ }\textbf {\bibinfo {volume} {53}},\ \bibinfo {pages}
  {473} (\bibinfo {year} {2001})}\BibitemShut {NoStop}%
\bibitem [{\citenamefont {Cassak}\ and\ \citenamefont
  {Shay}(2007)}]{cassak:asym}%
  \BibitemOpen
  \bibfield  {author} {\bibinfo {author} {\bibfnamefont {P.~A.}\ \bibnamefont
  {Cassak}}\ and\ \bibinfo {author} {\bibfnamefont {M.~A.}\ \bibnamefont
  {Shay}},\ }\href@noop {} {\bibfield  {journal} {\bibinfo  {journal} {Phys.
  Plasmas},\ }\textbf {\bibinfo {volume} {14}},\ \bibinfo {eid} {102114}
  (\bibinfo {year} {2007})}\BibitemShut {NoStop}%
\bibitem [{\citenamefont {Cassak}\ and\ \citenamefont
  {Shay}(2008)}]{cassak:hall}%
  \BibitemOpen
  \bibfield  {author} {\bibinfo {author} {\bibfnamefont {P.~A.}\ \bibnamefont
  {Cassak}}\ and\ \bibinfo {author} {\bibfnamefont {M.~A.}\ \bibnamefont
  {Shay}},\ }\href@noop {} {\bibfield  {journal} {\bibinfo  {journal} {Geophys.
  Res. Lett.},\ }\textbf {\bibinfo {volume} {35}},\ \bibinfo {eid} {19102}
  (\bibinfo {year} {2008})}\BibitemShut {NoStop}%
\bibitem [{\citenamefont {{La Belle-Hamer}}\ \emph {et~al.}(1995)\citenamefont
  {{La Belle-Hamer}}, \citenamefont {{Otto}},\ and\ \citenamefont
  {{Lee}}}]{labellehamer:1995}%
  \BibitemOpen
  \bibfield  {author} {\bibinfo {author} {\bibfnamefont {A.~L.}\ \bibnamefont
  {{La Belle-Hamer}}}, \bibinfo {author} {\bibfnamefont {A.}~\bibnamefont
  {{Otto}}}, \ and\ \bibinfo {author} {\bibfnamefont {L.~C.}\ \bibnamefont
  {{Lee}}},\ }\href@noop {} {\bibfield  {journal} {\bibinfo  {journal} {J.
  Geophys. Res.},\ }\textbf {\bibinfo {volume} {100}},\ \bibinfo {pages}
  {11875} (\bibinfo {year} {1995})}\BibitemShut {NoStop}%
\bibitem [{\citenamefont {{Ugai}}(2000)}]{ugai:2000}%
  \BibitemOpen
  \bibfield  {author} {\bibinfo {author} {\bibfnamefont {M.}~\bibnamefont
  {{Ugai}}},\ }\href@noop {} {\bibfield  {journal} {\bibinfo  {journal} {Phys.
  Plasmas},\ }\textbf {\bibinfo {volume} {7}},\ \bibinfo {pages} {867}
  (\bibinfo {year} {2000})}\BibitemShut {NoStop}%
\bibitem [{\citenamefont {{Kondoh}}\ \emph {et~al.}(2004)\citenamefont
  {{Kondoh}}, \citenamefont {{Ugai}},\ and\ \citenamefont
  {{Shimizu}}}]{kondoh:2004}%
  \BibitemOpen
  \bibfield  {author} {\bibinfo {author} {\bibfnamefont {K.}~\bibnamefont
  {{Kondoh}}}, \bibinfo {author} {\bibfnamefont {M.}~\bibnamefont {{Ugai}}}, \
  and\ \bibinfo {author} {\bibfnamefont {T.}~\bibnamefont {{Shimizu}}},\
  }\href@noop {} {\bibfield  {journal} {\bibinfo  {journal} {Adv. Space Res.},\
  }\textbf {\bibinfo {volume} {33}},\ \bibinfo {pages} {794} (\bibinfo {year}
  {2004})}\BibitemShut {NoStop}%
\bibitem [{\citenamefont {{Borovsky}}\ and\ \citenamefont
  {{Hesse}}(2007)}]{borovsky:asym}%
  \BibitemOpen
  \bibfield  {author} {\bibinfo {author} {\bibfnamefont {J.~E.}\ \bibnamefont
  {{Borovsky}}}\ and\ \bibinfo {author} {\bibfnamefont {M.}~\bibnamefont
  {{Hesse}}},\ }\href@noop {} {\bibfield  {journal} {\bibinfo  {journal} {Phys.
  Plasmas},\ }\textbf {\bibinfo {volume} {14}},\ \bibinfo {eid} {102309}
  (\bibinfo {year} {2007})}\BibitemShut {NoStop}%
\bibitem [{\citenamefont {{Birn}}\ \emph {et~al.}(2008)\citenamefont {{Birn}},
  \citenamefont {{Borovsky}},\ and\ \citenamefont {{Hesse}}}]{birn:2008}%
  \BibitemOpen
  \bibfield  {author} {\bibinfo {author} {\bibfnamefont {J.}~\bibnamefont
  {{Birn}}}, \bibinfo {author} {\bibfnamefont {J.~E.}\ \bibnamefont
  {{Borovsky}}}, \ and\ \bibinfo {author} {\bibfnamefont {M.}~\bibnamefont
  {{Hesse}}},\ }\href@noop {} {\bibfield  {journal} {\bibinfo  {journal} {Phys.
  Plasmas},\ }\textbf {\bibinfo {volume} {15}},\ \bibinfo {eid} {032101}
  (\bibinfo {year} {2008})}\BibitemShut {NoStop}%
\bibitem [{\citenamefont {{Pritchett}}(2008)}]{pritchett:2008}%
  \BibitemOpen
  \bibfield  {author} {\bibinfo {author} {\bibfnamefont {P.~L.}\ \bibnamefont
  {{Pritchett}}},\ }\href@noop {} {\bibfield  {journal} {\bibinfo  {journal}
  {J. Geophys. Res.},\ }\textbf {\bibinfo {volume} {113}},\ \bibinfo {pages}
  {6210} (\bibinfo {year} {2008})}\BibitemShut {NoStop}%
\bibitem [{\citenamefont {{Birn}}\ \emph {et~al.}(2010)\citenamefont {{Birn}},
  \citenamefont {{Borovsky}}, \citenamefont {{Hesse}},\ and\ \citenamefont
  {{Schindler}}}]{birn:2010}%
  \BibitemOpen
  \bibfield  {author} {\bibinfo {author} {\bibfnamefont {J.}~\bibnamefont
  {{Birn}}}, \bibinfo {author} {\bibfnamefont {J.~E.}\ \bibnamefont
  {{Borovsky}}}, \bibinfo {author} {\bibfnamefont {M.}~\bibnamefont {{Hesse}}},
  \ and\ \bibinfo {author} {\bibfnamefont {K.}~\bibnamefont {{Schindler}}},\
  }\href@noop {} {\bibfield  {journal} {\bibinfo  {journal} {Phys. Plasmas},\
  }\textbf {\bibinfo {volume} {17}},\ \bibinfo {eid} {052108} (\bibinfo {year}
  {2010})}\BibitemShut {NoStop}%
\bibitem [{\citenamefont {{Bemporad}}\ \emph {et~al.}(2007)\citenamefont
  {{Bemporad}}, \citenamefont {{Raymond}}, \citenamefont {{Poletto}},\ and\
  \citenamefont {{Romoli}}}]{bemporad:2007}%
  \BibitemOpen
  \bibfield  {author} {\bibinfo {author} {\bibfnamefont {A.}~\bibnamefont
  {{Bemporad}}}, \bibinfo {author} {\bibfnamefont {J.}~\bibnamefont
  {{Raymond}}}, \bibinfo {author} {\bibfnamefont {G.}~\bibnamefont
  {{Poletto}}}, \ and\ \bibinfo {author} {\bibfnamefont {M.}~\bibnamefont
  {{Romoli}}},\ }\href@noop {} {\bibfield  {journal} {\bibinfo  {journal}
  {Astrophys. J.},\ }\textbf {\bibinfo {volume} {655}},\ \bibinfo {pages} {576}
  (\bibinfo {year} {2007})}\BibitemShut {NoStop}%
\bibitem [{\citenamefont {Vourlidas}\ \emph {et~al.}(2010)\citenamefont
  {Vourlidas}, \citenamefont {Howard}, \citenamefont {Esfandiari},
  \citenamefont {Patsourakos}, \citenamefont {Yashiro},\ and\ \citenamefont
  {Michalek}}]{vourlidas:2010}%
  \BibitemOpen
  \bibfield  {author} {\bibinfo {author} {\bibfnamefont {A.}~\bibnamefont
  {Vourlidas}}, \bibinfo {author} {\bibfnamefont {R.~A.}\ \bibnamefont
  {Howard}}, \bibinfo {author} {\bibfnamefont {E.}~\bibnamefont {Esfandiari}},
  \bibinfo {author} {\bibfnamefont {S.}~\bibnamefont {Patsourakos}}, \bibinfo
  {author} {\bibfnamefont {S.}~\bibnamefont {Yashiro}}, \ and\ \bibinfo
  {author} {\bibfnamefont {G.}~\bibnamefont {Michalek}},\ }\bibfield  {title}
  {\enquote {\bibinfo {title} {Comprehensive analysis of coronal mass ejection
  mass and energy properties over a full solar cycle},}\ }\href@noop {}
  {\bibfield  {journal} {\bibinfo  {journal} {Astrophys.\ J., in press}}
  (\bibinfo {year} {2010})}\BibitemShut {NoStop}%
\bibitem [{\citenamefont {{Lin}}\ \emph {et~al.}(2004)\citenamefont {{Lin}},
  \citenamefont {{Raymond}},\ and\ \citenamefont {{van
  Ballegooijen}}}]{lin:2004}%
  \BibitemOpen
  \bibfield  {author} {\bibinfo {author} {\bibfnamefont {J.}~\bibnamefont
  {{Lin}}}, \bibinfo {author} {\bibfnamefont {J.~C.}\ \bibnamefont
  {{Raymond}}}, \ and\ \bibinfo {author} {\bibfnamefont {A.~A.}\ \bibnamefont
  {{van Ballegooijen}}},\ }\href@noop {} {\bibfield  {journal} {\bibinfo
  {journal} {Astrophys. J.},\ }\textbf {\bibinfo {volume} {602}},\ \bibinfo
  {pages} {422} (\bibinfo {year} {2004})}\BibitemShut {NoStop}%
\bibitem [{\citenamefont {{Akmal}}\ \emph {et~al.}(2001)\citenamefont
  {{Akmal}}, \citenamefont {{Raymond}}, \citenamefont {{Vourlidas}},
  \citenamefont {{Thompson}}, \citenamefont {{Ciaravella}}, \citenamefont
  {{Ko}}, \citenamefont {{Uzzo}},\ and\ \citenamefont {{Wu}}}]{akmal:2001}%
  \BibitemOpen
  \bibfield  {author} {\bibinfo {author} {\bibfnamefont {A.}~\bibnamefont
  {{Akmal}}}, \bibinfo {author} {\bibfnamefont {J.~C.}\ \bibnamefont
  {{Raymond}}}, \bibinfo {author} {\bibfnamefont {A.}~\bibnamefont
  {{Vourlidas}}}, \bibinfo {author} {\bibfnamefont {B.}~\bibnamefont
  {{Thompson}}}, \bibinfo {author} {\bibfnamefont {A.}~\bibnamefont
  {{Ciaravella}}}, \bibinfo {author} {\bibfnamefont {Y.-K.}\ \bibnamefont
  {{Ko}}}, \bibinfo {author} {\bibfnamefont {M.}~\bibnamefont {{Uzzo}}}, \ and\
  \bibinfo {author} {\bibfnamefont {R.}~\bibnamefont {{Wu}}},\ }\href@noop {}
  {\bibfield  {journal} {\bibinfo  {journal} {Astrophys. J.},\ }\textbf
  {\bibinfo {volume} {553}},\ \bibinfo {pages} {922} (\bibinfo {year}
  {2001})}\BibitemShut {NoStop}%
\bibitem [{\citenamefont {{Ciaravella}}\ \emph {et~al.}(2001)\citenamefont
  {{Ciaravella}}, \citenamefont {{Raymond}}, \citenamefont {{Reale}},
  \citenamefont {{Strachan}},\ and\ \citenamefont {{Peres}}}]{ciaravella:2001}%
  \BibitemOpen
  \bibfield  {author} {\bibinfo {author} {\bibfnamefont {A.}~\bibnamefont
  {{Ciaravella}}}, \bibinfo {author} {\bibfnamefont {J.~C.}\ \bibnamefont
  {{Raymond}}}, \bibinfo {author} {\bibfnamefont {F.}~\bibnamefont {{Reale}}},
  \bibinfo {author} {\bibfnamefont {L.}~\bibnamefont {{Strachan}}}, \ and\
  \bibinfo {author} {\bibfnamefont {G.}~\bibnamefont {{Peres}}},\ }\href@noop
  {} {\bibfield  {journal} {\bibinfo  {journal} {Astrophys. J.},\ }\textbf
  {\bibinfo {volume} {557}},\ \bibinfo {pages} {351} (\bibinfo {year}
  {2001})}\BibitemShut {NoStop}%
\bibitem [{\citenamefont {{Lee}}\ \emph {et~al.}(2009)\citenamefont {{Lee}},
  \citenamefont {{Raymond}}, \citenamefont {{Ko}},\ and\ \citenamefont
  {{Kim}}}]{lee:2009}%
  \BibitemOpen
  \bibfield  {author} {\bibinfo {author} {\bibfnamefont {J.-Y.}\ \bibnamefont
  {{Lee}}}, \bibinfo {author} {\bibfnamefont {J.~C.}\ \bibnamefont
  {{Raymond}}}, \bibinfo {author} {\bibfnamefont {Y.-K.}\ \bibnamefont {{Ko}}},
  \ and\ \bibinfo {author} {\bibfnamefont {K.-S.}\ \bibnamefont {{Kim}}},\
  }\href@noop {} {\bibfield  {journal} {\bibinfo  {journal} {Astrophys. J.},\
  }\textbf {\bibinfo {volume} {692}},\ \bibinfo {pages} {1271} (\bibinfo {year}
  {2009})}\BibitemShut {NoStop}%
\bibitem [{\citenamefont {{Landi}}\ \emph {et~al.}(2010)\citenamefont
  {{Landi}}, \citenamefont {{Raymond}}, \citenamefont {{Miralles}},\ and\
  \citenamefont {{Hara}}}]{landi:2010}%
  \BibitemOpen
  \bibfield  {author} {\bibinfo {author} {\bibfnamefont {E.}~\bibnamefont
  {{Landi}}}, \bibinfo {author} {\bibfnamefont {J.~C.}\ \bibnamefont
  {{Raymond}}}, \bibinfo {author} {\bibfnamefont {M.~P.}\ \bibnamefont
  {{Miralles}}}, \ and\ \bibinfo {author} {\bibfnamefont {H.}~\bibnamefont
  {{Hara}}},\ }\href@noop {} {\bibfield  {journal} {\bibinfo  {journal}
  {Astrophys. J.},\ }\textbf {\bibinfo {volume} {711}},\ \bibinfo {pages} {75}
  (\bibinfo {year} {2010})}\BibitemShut {NoStop}%
\bibitem [{\citenamefont {{Furth}}\ \emph {et~al.}(1963)\citenamefont
  {{Furth}}, \citenamefont {{Killeen}},\ and\ \citenamefont
  {{Rosenbluth}}}]{fkr}%
  \BibitemOpen
  \bibfield  {author} {\bibinfo {author} {\bibfnamefont {H.~P.}\ \bibnamefont
  {{Furth}}}, \bibinfo {author} {\bibfnamefont {J.}~\bibnamefont {{Killeen}}},
  \ and\ \bibinfo {author} {\bibfnamefont {M.~N.}\ \bibnamefont
  {{Rosenbluth}}},\ }\href@noop {} {\bibfield  {journal} {\bibinfo  {journal}
  {Phys. Fluids},\ }\textbf {\bibinfo {volume} {6}},\ \bibinfo {pages} {459}
  (\bibinfo {year} {1963})}\BibitemShut {NoStop}%
\bibitem [{\citenamefont {{Lin}}\ \emph {et~al.}(2007)\citenamefont {{Lin}},
  \citenamefont {{Li}}, \citenamefont {{Forbes}}, \citenamefont {{Ko}},
  \citenamefont {{Raymond}},\ and\ \citenamefont
  {{Vourlidas}}}]{2007ApJ...658L.123L}%
  \BibitemOpen
  \bibfield  {author} {\bibinfo {author} {\bibfnamefont {J.}~\bibnamefont
  {{Lin}}}, \bibinfo {author} {\bibfnamefont {J.}~\bibnamefont {{Li}}},
  \bibinfo {author} {\bibfnamefont {T.~G.}\ \bibnamefont {{Forbes}}}, \bibinfo
  {author} {\bibfnamefont {Y.}~\bibnamefont {{Ko}}}, \bibinfo {author}
  {\bibfnamefont {J.~C.}\ \bibnamefont {{Raymond}}}, \ and\ \bibinfo {author}
  {\bibfnamefont {A.}~\bibnamefont {{Vourlidas}}},\ }\href@noop {} {\bibfield
  {journal} {\bibinfo  {journal} {Astrophys. J. Lett.},\ }\textbf {\bibinfo
  {volume} {658}},\ \bibinfo {pages} {L123} (\bibinfo {year}
  {2007})}\BibitemShut {NoStop}%
\bibitem [{\citenamefont {{Lazarian}}\ and\ \citenamefont
  {{Vishniac}}(1999)}]{lazarian:turbrecon}%
  \BibitemOpen
  \bibfield  {author} {\bibinfo {author} {\bibfnamefont {A.}~\bibnamefont
  {{Lazarian}}}\ and\ \bibinfo {author} {\bibfnamefont {E.~T.}\ \bibnamefont
  {{Vishniac}}},\ }\href@noop {} {\bibfield  {journal} {\bibinfo  {journal}
  {Astrophys. J.},\ }\textbf {\bibinfo {volume} {517}},\ \bibinfo {pages} {700}
  (\bibinfo {year} {1999})}\BibitemShut {NoStop}%
\bibitem [{\citenamefont {Oka}(2010)}]{oka:pc:2010}%
  \BibitemOpen
  \bibfield  {author} {\bibinfo {author} {\bibfnamefont {M.}~\bibnamefont
  {Oka}},\ }\href@noop {} {} (\bibinfo {year} {2010}),\ \bibinfo {note}
  {private communication}\BibitemShut {NoStop}%
\end{thebibliography}


%


\end{document}